\documentclass{wscpaperproc}

\usepackage{amsmath}
\usepackage{amsfonts}
\usepackage{amssymb}
\usepackage{amsbsy}
\usepackage{amsthm}
\usepackage[pdftex,colorlinks=true,urlcolor=blue,citecolor=black,anchorcolor=black,linkcolor=black]{hyperref}

\usepackage{microtype}
\usepackage{graphicx}
\usepackage{subfigure}
\usepackage{booktabs} % for professional tables
\usepackage[ruled,vlined]{algorithm2e}
\usepackage{latexsym}
\usepackage{graphicx}
\usepackage{mathptmx}
\usepackage[T1]{fontenc}

\usepackage[utf8]{inputenc}
\usepackage{multirow}
\usepackage{url}            % simple URL typesetting
\usepackage{nicefrac}       % compact symbols for 1/2, etc.
\usepackage{bbold}
\usepackage{capt-of}
\usepackage{etoc}
\usepackage{caption}

\usepackage{tabularx,makecell}

\usepackage{epsfig,epstopdf,color,soul}
\allowdisplaybreaks
\usepackage{color}
\usepackage{enumerate}
\usepackage[shortlabels]{enumitem}
\usepackage{multicol}
\usepackage{empheq}
\usepackage{caption} 
\newcommand{\Bcal}{{\mathcal B}}

\newcommand{\Dcal}{{\mathcal D}}

\newcommand{\Ocal}{{\mathcal O}}

\newcommand{\Scal}{{\mathcal S}}

\newcommand{\Vcal}{{\mathcal V}}

\newcommand{\1}{{\mathbf{1}}}

\newcommand{\argmin}{\mathop{\rm argmin}}

\newtheorem{lem}{Lemma}
\newtheorem{thm}{Theorem}

\newtheorem{definition}{Definition}

\newcommand{\NSW}{\operatorname{\mathbf{NSW}}}
\newcommand{\logNSW}{\operatorname{\mathbf{logNSW}}}
\newcommand{\exploitability}{\operatorname{\mathbf{expl}}}
\newcommand{\efficiency}{\operatorname{\mathbf{efficiency}}}

\usepackage{tcolorbox}
\tcbuselibrary{theorems}

\newtcbtheorem{mechanism}{Mechanism}%
{}{mec}

\newtheoremstyle{wsc}% hnamei
{3pt}% hSpace abovei
{3pt}% hSpace belowi
{}% hBody fonti
{}% hIndent amounti1
{\bf}% hTheorem head fontbf
{}% hPunctuation after theorem headi
{.5em}% hSpace after theorem headi2
{}% hTheorem head spec (can be left empty, meaning `normal')i

\theoremstyle{wsc}
\newtheorem{theorem}{Theorem}

\newtheorem{proposition}{Proposition}

\begin{document}

%***************************************************************************
% AUTHOR: AUTHOR NAMES GO HERE
% FORMAT AUTHORS NAMES Like: Author1, Author2 and Author3 (last names)
%
%		You need to change the author listing below!
%               Please list ALL authors using last name only, separate by a comma except
%               for the last author, separate with "and"
%

% setting up general page style
\pagestyle{fancyplain}

% setting up page style of first page
\thispagestyle{plain}
\firstPageHead{}

% setting up running header (authors) of subsequent pages
\chead{\fancyplain{}{\itshape Zeng, Bhatt, Kreacic, Hassanzadeh, Koppel, and Ganesh}}

% setting up seperation parameters
%\headsep=72pt
\rhead{}
\cfoot{}
\renewcommand{\headrulewidth}{0pt} % (renewcommand needed in fancyhdr to remove top decorative line)
%\headrulewidth=0pt  % ("setlength" needed in fancyheading to remove top decorative line)

%%%%%%%%%%%%%%%%%%%%%%%%%%%%%%%%%%%%%%%%%%%%%%%%%%%%%%%%%%%%%%%%%%%%%%%%%%%%%%
%                                                                            %
%     THESE COMMANDS ARE REQUIRED TO WORK WITH WSC.BST TO MAKE BIBLIO     %
%                                                                            %
%%%%%%%%%%%%%%%%%%%%%%%%%%%%%%%%%%%%%%%%%%%%%%%%%%%%%%%%%%%%%%%%%%%%%%%%%%%%%%
\makeatletter
\let\@internalcite\cite
\def\cite{\def\@citeseppen{-1000}%
    \def\@cite##1##2{(##1\if@tempswa , ##2\fi)}%
    \def\citeauthoryear##1##2##3{##1 ##3}\@internalcite}
\def\citeNP{\def\@citeseppen{-1000}%
    \def\@cite##1##2{##1\if@tempswa , ##2\fi}%
    \def\citeauthoryear##1##2##3{##1 ##3}\@internalcite}
\def\citeN{\def\@citeseppen{-1000}%
%  Pierre L'Ecuyer's fix for multiple cite bug
%  Added by Paul J Sanchez on 4 October 2001
%   \def\@cite##1##2{##1\if@tempswa , ##2)\else{)}\fi}%
%   \def\citeauthoryear##1##2##3{##1 (##3}\@citedata}
    \def\@cite##1##2{##1\if@tempswa, ##2)\else{}\fi}%
    \def\citeauthoryear##1##2##3{##1 (##3)}\@citedata}
\def\citeA{\def\@citeseppen{-1000}%
    \def\@cite##1##2{(##1\if@tempswa , ##2\fi)}%
    \def\citeauthoryear##1##2##3{##1}\@internalcite}
\def\citeANP{\def\@citeseppen{-1000}%
    \def\@cite##1##2{##1\if@tempswa , ##2\fi}%
    \def\citeauthoryear##1##2##3{##1}\@internalcite}
\def\shortcite{\def\@citeseppen{-1000}%
    \def\@cite##1##2{(##1\if@tempswa , ##2\fi)}%
    \def\citeauthoryear##1##2##3{##2 ##3}\@internalcite}
\def\shortciteNP{\def\@citeseppen{-1000}%
    \def\@cite##1##2{##1\if@tempswa , ##2\fi}%
    \def\citeauthoryear##1##2##3{##2 ##3}\@internalcite}
\def\shortciteN{\def\@citeseppen{-1000}%
%  Pierre L'Ecuyer's fix for multiple cite bug
%  Added by Paul J Sanchez on 2 September 2002
%  should have caught this last year...
%   \def\@cite##1##2{##1\if@tempswa , ##2)\else{)}\fi}%
%   \def\citeauthoryear##1##2##3{##2 (##3}\@citedata}
% Shane G. Henderson fix for extra right bracket at end of optional material June 8, 2005
%    \def\@cite##1##2{##1\if@tempswa, ##2)\else{}\fi}%
    \def\@cite##1##2{##1\if@tempswa, ##2\else{}\fi}%
    \def\citeauthoryear##1##2##3{##2 (##3)}\@citedata}
\def\shortciteA{\def\@citeseppen{-1000}%
    \def\@cite##1##2{(##1\if@tempswa , ##2\fi)}%
    \def\citeauthoryear##1##2##3{##2}\@internalcite}
\def\shortciteANP{\def\@citeseppen{-1000}%
    \def\@cite##1##2{##1\if@tempswa , ##2\fi}%
    \def\citeauthoryear##1##2##3{##2}\@internalcite}
\def\citeyear{\def\@citeseppen{-1000}%
    \def\@cite##1##2{(##1\if@tempswa , ##2\fi)}%
    \def\citeauthoryear##1##2##3{##3}\@citedata}
\def\citeyearNP{\def\@citeseppen{-1000}%
    \def\@cite##1##2{##1\if@tempswa , ##2\fi}%
    \def\citeauthoryear##1##2##3{##3}\@citedata}
%
% \@citedata and \@citedatax:
%
% Place commas in-between citations in the same \citeyear, \citeyearNP,
% \citeN, or \shortciteN command.
% Use something like \citeN{ref1,ref2,ref3} and \citeN{ref4} for a list.
%
\def\@citedata{%
    \@ifnextchar [{\@tempswatrue\@citedatax}%
                  {\@tempswafalse\@citedatax[]}%
}

\def\@citedatax[#1]#2{%
\if@filesw\immediate\write\@auxout{\string\citation{#2}}\fi%
  \def\@citea{}\@cite{\@for\@citeb:=#2\do%
    {\@citea\def\@citea{, }\@ifundefined% by Young
       {b@\@citeb}{{\bf ?}%
       \@warning{Citation `\@citeb' on page \thepage \space undefined}}%
{\csname b@\@citeb\endcsname}}}{#1}}%

% don't box citations, separate with ; and a space
% also, make the penalty between citations negative: a good place to break.
%
\def\@citex[#1]#2{%
\if@filesw\immediate\write\@auxout{\string\citation{#2}}\fi%
  \def\@citea{}\@cite{\@for\@citeb:=#2\do%
    {\@citea\def\@citea{; }\@ifundefined% by Young
       {b@\@citeb}{{\bf ?}%
       \@warning{Citation `\@citeb' on page \thepage \space undefined}}%
{\csname b@\@citeb\endcsname}}}{#1}}%

% (from apalike.sty)
% No labels in the bibliography.
%
\def\@biblabel#1{}
\makeatother

%\newlength{\bibhang}
%\setlength{\bibhang}{2em}

% Indent second and subsequent lines of bibliographic entries. Taken
% from openbib.sty: \newblock is set to {}.
% \renewcommand{\refname}{REFERENCES}

\newdimen\bibindent
\bibindent=0.0em
% SEC: was \def\thebibliography#1{\section*{\refname\@mkboth
% SEC: was   {\uppercase{\refname}}{\uppercase{\refname}}}\list
\def\thebibliography#1{\section*{\refname}\list
   {}{\settowidth\labelwidth{[#1]}
   \leftmargin\parindent
   \itemindent -\parindent
   \listparindent \itemindent
   \itemsep 0pt
   \parsep 0pt}
   \def\newblock{}
   \sloppy
   \sfcode`\.=1000\relax}

           % Set up BiBTeX macros

% needed to make the tex document look more like the word counterpart :-(
\setlength{\baselineskip}{12.7pt}

% AUTHOR: Enter the title, all letters in upper case
% \title{Neural Learning for Payment-free Resource Allocation}
\title{Learning Payment-Free Resource Allocation Mechanisms}

% AUTHOR: Enter the authors of the article, see end of the example document for further examples
\author{\begin{center}Sihan Zeng\textsuperscript{1}, Sujay Bhatt\textsuperscript{1},  Eleonora Kreacic\textsuperscript{1}, Parisa Hassanzadeh\textsuperscript{2}, Alec Koppel\textsuperscript{1}, and Sumitra Ganesh\textsuperscript{1}\\
[11pt]
\textsuperscript{1}J.P. Morgan AI Research, USA \& UK\qquad\textsuperscript{2}Samsung Data Systems, USA\end{center}
}

\maketitle

\vspace{-12pt}

\section*{ABSTRACT}
We consider the design of mechanisms that allocate limited resources among self-interested agents using neural networks. Unlike the recent works that leverage machine learning for revenue maximization in auctions, we consider welfare maximization as the key objective in the \textit{payment-free} setting. 
% In this setting, unlike auctions, 
Without payment exchange,
it is unclear how we can align agents' incentives to achieve the desired objectives of truthfulness and social welfare simultaneously, without resorting to approximations. Our work makes novel contributions by designing an \emph{approximate} mechanism that desirably trade-off social welfare with truthfulness. Specifically, (i)~we contribute a new end-to-end neural network architecture, \texttt{ExS-Net}, that accommodates the idea of ``money-burning'' for mechanism design without payments;
(ii)~we provide a generalization bound that guarantees the mechanism performance when trained under finite samples; 
and (iii) we provide an experimental demonstration of the merits of the proposed mechanism.

%!TEX root = main.tex
%
\section{Introduction}\label{sec:intro}
%\textcolor{blue}{Tentative paper outline: 
%\begin{itemize}
%    \item Introduction. Related Work. Discussion of Dutting et.al. Motivation.
%    \item Model \& Problem setup. Metrics with motivation.
%    \item Learning RA without money:\\ (a) Money-burning \\
%    (b) Architecture design
%    \item Baseline algorithms: PF \& PA \\
%    (a) Sweeping for small cases. \\
%    (b) Say large cases unified treatment .. future work.
%    \item Experiments \\    
%    (a) Realistic data: Motivate a RA problem \\
%    (b) Synthetic data. small systems -- Insights using heat-map.
%\end{itemize}
%}
Mechanism design studies how to induce a game among strategic agents in such a way that the induced game satisfies a set of desired properties at the equilibrium. These properties include individually rationality (agents are motivated to participate in the game), incentive compatibility (agents are motivated to report private information truthfully), social welfare (socially desirable outcome is chosen), efficiency (resources are not wasted when there is demand), envy-freeness (agents do not wish for others' share of the resources); to name a few. 
% It is well known in mechanism design that these properties cannot be simultaneously satisfied exactly by any given mechanism~\shortcite{nisan2007introduction,borgers2015introduction,cole2013mechanism}. 
In this work, we are interested in \textit{incentive compatibility}, \textit{social welfare} and \textit{efficiency}; and we propose learning the rules of the game so that the agents are motivated to report their preferences truthfully in a \textit{payment-free setting}, and a socially desirable outcome that is fair and efficient results in the equilibrium.   

\subsection{Main Contribution}
We consider multiple-agents and multiple divisible items{\footnote{A crude way to handle the indivisible setting using our framework is via rounding~\shortcite{cole2015approximating}, where the agent receiving the highest fraction is allocated the item.}}, and aim to design an incentive compatible (\textbf{IC}) mechanism that maximizes social welfare \& efficiency in the payments-free setting. Since the three objectives cannot be achieved simultaneously as shown in ~\shortciteN{cole2013mechanism}, we consider learning an approximate mechanism that achieve a desirable trade-off. To provide an overview of our work:
\begin{itemize}
    \item We consider Nash Social Welfare (NSW) as the welfare objective for the mechanism designer. We further impose constraints on the approximate incentive compatibility of the mechanism. These constraints are defined using the notion of ``exploitability''~\shortcite{GG22} -- which measures the maximum utility gain when agents deviate from truthfully reporting. 
    \item We design a novel ``state-augmentation with an artificial agent'' based neural architecture -- named \texttt{ExS-Net} -- that simulates ``money-burning''~\shortcite{HR08} in the ``hardware'', i.e., intentional withholding of resources as an implicit form of payment. In addition, we make the design modular, in that, any neural network (feed-forward, CNN, Transformers) can be used as the hidden layer and the money-burning hardware only modifies the output/ activation layer. 
    \item We derive the generalization bounds for \texttt{Ex-Net} and also provide guarantees under distribution shift. The robustness to distribution shift in the experiments, specifically, demonstrates the significance of our contribution by allowing the training data itself to be subjected to adversarial contamination.
    \item Extensive experiments confirm the efficacy of the proposed mechanisms in achieving the desired trade-off. We explicitly evaluate the exploitability and social welfare of our mechanism compared to the baseline mechanisms Proportionally Fair (PF) and Partial Allocation (PA)~\shortcite{cole2013mechanism}. The proposed mechanism~\texttt{ExS-Net} significantly outperforms PF in terms of exploitability and PA in terms of efficiency \& social welfare. 
\end{itemize}

% \subsection{Motivating Examples
% % from Business
% }
% We include a few JP Morgan business use cases that motivate the setting considered in this paper{\footnote{Owing to the sensitive nature of the associated confidential data, we report experiments on general distributions.}}. Key aspects include welfare maximizing, efficient, and IC resource allocation without the payments.
% \begin{itemize}
%     \item \textbf{Computing}: Firm allocates budget constrained computing resources (memory, GPUs) to various internal business and research teams for assisting in their individual  goals. Each team is self-interested and could potentially misreport the preferences and demand to gain increased allocations. A common occurrence is when the research team is facing an impending conference deadline. Designing fair allocation mechanisms that ensures that the teams have little incentive to report untruthfully and are efficient, are necessary for cost reduction and improving overall satisfaction.
%     \item \textbf{Auto-loans}: Self-interested customers request for loan packages from the firm via dealers and in-turn share experiences on third-party trusted websites. Dealers act as intermediaries for vetting the customer application and bringing in the business. Designing mechanisms that secure fair \& efficient loan approvals for customers while ensuring they share truthful account of their experiences on the platform is important in achieving AI for social good.
% \end{itemize}

\subsection{Related Literature}
We organize the literature review into the different resource allocation problem classes, relevant solution approaches, and the key differences from the most relevant work.\\ %While highlighting how our work is placed in the literature, we also provide information on the baselines that are compared in the main contributions. \\

\vspace{-10pt}
\noindent \textbf{\textit{Resource Allocation Settings}}\\
Classic work on resource allocation focuses on divisible resources, with recent interest spanning both the indivisible and divisible setting.
In either case, payment is a frequently used tool to align incentives~\shortcite{pavlov2011optimal,giannakopoulos2014duality,dütting2022optimal} and ensure the truthful behavior of resource consumers. However, payment is naturally forbidden in many settings including organ donation, food and necessity distribution by charity, allocation of GPU hours by an institution to its employees \shortcite{DFP10,PT13}. Our work focuses on the divisible setting with multiple-agents and items, and considers designing a fair \& efficient resource allocation mechanism that is incentive compatible without payments. Since the three objectives cannot be achieved simultaneously, we consider learning approximate mechanisms that achieves a desirable trade-off.  \\

\vspace{-10pt}
\noindent \textbf{\textit{Solution Approaches for Payment-Free Setting}} \\
% Proportional Fairness (PF) was first defined in~\shortciteN{Key97} in the context of TCP congestion control. This naturally defines a mechanism that allocates resources in a fair and efficient manner among agents. PF is the most well-known (first) solution as it is known to be equivalent to Competitive Equilibria with Equal Outcomes (CEEI)~\shortcite{Var73} and also to the widely-known Nash bargaining solution~\shortcite{nash1950bargaining}, and most importantly maximizes Nash Social Welfare~\shortcite{bertsimas2011price}; the main objective in our work.  
Ensuring allocation is fair among agents may be formalized through \textit{proportional fairness} (\textbf{PF}) \shortcite{Key97}. PF allocations are such that if an alternate allocation is adopted, the percentage utility gain over all agents sums up to a non-positive number. Importantly, PF is achieved by maximizing Nash social welfare (\textbf{NSW}), the product of all agents' utilities \shortcite{bertsimas2011price}, which we use as a quantifier of social welfare in this work. The PF mechanism is one that directly seeks to optimize NSW by solving a constrained optimization program, and by definition achieves the maximum possible NSW under allocation constraints.

% \shortciteN{cole2013mechanism} identifies a key issue with PF -- it \textit{cannot} be achieved by truthful mechanisms. 
% In fact, it is proved that no IC mechanism can guarantee to every agent an allocation larger than one half. In other words, only a fraction of the valuation that the agent deserves according to the PF solution is allocated when aligning it's preferences truthfully. 
\shortciteN{cole2013mechanism} identifies a key issue with the PF mechanism -- strategic agents may misreport their utility functions to gain inflated allocation from a supplier running the PF mechanism. 
% In fact, it is proved that no IC mechanism can guarantee to every agent an allocation larger than one half. In other words, only a fraction of the valuation that the agent deserves according to the PF solution is allocated when aligning it's preferences truthfully. 
To prevent such untruthful behavior, \shortciteN{cole2013mechanism} design a novel mechanism -- partial allocation (\textbf{PA}) -- which achieves the exact IC at the cost of efficiency and NSW. Prior to our work, PA is the state-of-the-art in balancing NSW and incentive compatibility. The main contribution of our work is to achieve an even more desirable operating point between NSW and incentive compatibility. \\

\vspace{-10pt}
% \noindent \textbf{\textit{Key Deviations from Neural Learning Literature}} \\
\noindent \textbf{\textit{Learning-Based Mechanism Design in Resource Allocation}} \\
In the literature on mechanism design for resource allocation, key gaps remain due to the difficulty of hand-designing approximate mechanisms that are ``somewhere in-between'' the desired objectives. For example, the PF mechanism optimizes NSW but is not IC; PA is hand-design to achieve IC but suffers sub-optimal NSW. 
Recently, there has been efforts to take a data-driven approach to resource allocation using neural networks. \shortciteN{dütting2022optimal} leads the way by learning payment-based allocation mechanism that not only approximate well-known optimal solutions for special cases, but also demonstrate how the solutions look like for multi-item \& multi-agent settings where the optimal solution is unknown. 
Our work deviates from the 
existing literature as we have to design IC mechanisms in the more challenging payment-free setting. The only known approach to align the incentive of the agents with that of the supplier without payment is ``money-burning''~\shortcite{HR08} -- intentional withholding of resources as an implicit form of payment. Inspired by this domain knowledge, we design a neural-network-based mechanism that approximates ``money-burning''.

% above neural learning literature in the following way: Unlike auctions that maximize revenue from payments, our work requires preference alignment without payments while maximizing welfare. The only known solution to align preferences without payment requires ``money-burning''~\shortcite{HR08} -- intentional withholding of resources as an implicit form of payment. We design a novel `state-augmentation with an artificial agent' based NN architecture that simulates money-burning in the `hardware'. In addition, we make the design modular, in that, any neural network (feed-forward, CNN, Transformers) can be used as the hidden layer and the money-burning hardware only modifies the output/ activation layer. 

\section{Mechanism Design without Payments}\label{sec:problem}
% A supplier allocates a finite number $M$ of divisible resources to  to $N$ agents. Let $a_{i,m} \in \mathbb{R}^{N M}$ denote the amount of resource $m$ allocated to agent $i$. There is a budget $b_m\geq 0$ on each resource $m=1,\cdots,M$, which specifies the amount of available resources. The value and demand for each resource is measured using scalars~$v_{i,m} \geq 0$ and $x_{i,m} \geq 0$ respectively. We use $\Acal_{b,x}$ to denote the set of valid allocations under budget $b=\{b_m\}_{m=1,\cdots,M}$ and demand $x$ which needs to satisfy $\Acal_{b,x}=\{a\in\mathbb{R}^{N M}:0\leq a_{i,m}\leq x_{i,m}, \forall i,m,\,\sum_{i=1}^{N}a_{i,m}\leq b_i,\forall m\}$.

% \subsection{Model \& Preliminaries}
% Every agent has an additive linear utility -- each unit of resource  increases the utility of agent by its value up to the demand threshold. In particular,  the utility{\footnote{We note that the mechanisms proposed in the paper could easily handle alternative utilities, as it is a learning based approach.}} function of each agent~$i$ is given by
% \begin{align}
%     u_i(a,v,x) \triangleq {\textstyle\sum}_{m=1}^{M}v_{i,m}\min\{a_{i,m},x_{i,m}\}. \label{def:utility}
% \end{align}
% for any $v\in\Vcal$ and $x\in\Dcal$. 

A supplier allocates a finite number $M$ of divisible resources to $N$ agents. The allocation can be represented as a vector $a\in\mathbb{R}^{NM}$, where $a_{i,m}$ represents the quantity of resource $m\in[M]$ allocated to agent $n\in[N]$. The supplier observes the budget $b_m\geq 0$ on each resource $m$, which specifies the quantity of the resource available. We denote $b=[b_1,\cdots,b_M]^{\top}\in\Bcal\subseteq\mathbb{R}_+^{M}$, where $\Bcal$ is the set of all possible budgets.
The agent's satisfaction with an allocation outcome is measured by a utility function. In this work, we assume that every agent has a (threshold) additive linear utility -- each unit of resource $m$ increases the utility of agent $i$ by value $v_{i,m}$ up to the threshold $x_{i,m}$ which represents demand. We aggregate valuations and demands by defining $v_i=[v_{i,1},\cdots,v_{i,M}]^{\top}$, $x_i=[x_{i,1},\cdots,x_{i,M}]^{\top}$, and $v=[v_1^{\top},\cdots,v_N^{\top}]^{\top}$, $x=[x_1^{\top},\cdots,x_N^{\top}]^{\top}$. These quantities are in respective spaces of values and demand $\Vcal\subseteq[\underline{v},\overline{v}]^{NM}$, $\Dcal\subseteq[\underline{d},\overline{d}]^{NM}$ for some scalars $\underline{v},\overline{v}>0$,  $\underline{d},\overline{d}\geq0$.
Given an allocation outcome $a\in\mathbb{R}^{NM}$, the utility function of agent $i$ is $u_i:\mathbb{R}_+^{NM}\times\Vcal\times\Dcal\rightarrow\mathbb{R}_+$ expressed as
\begin{align}
u_i(a,v,x) \triangleq {\textstyle\sum}_{m=1}^{M}v_{i,m}\min\{a_{i,m},x_{i,m}\}. \label{def:utility}
\end{align}

This utility models satisfaction in real-life problems and is widely studied \shortcite{bliem2016complexity,camacho2021beyond}. 
Our work operates in the common setting where the supplier knows the functional form of the utility function and relies on each agent $i$ to report its parameters $v_{i,m}$ and $x_{i,m}$ \shortcite{PT13,cole2013mechanism,dütting2022optimal}. 
% \begin{definition}[Mechanism]
% % \label{def:mechanism}
%     A mapping $f:\Vcal\times\Dcal\times\Bcal\rightarrow\mathbb{R}_+^{NM}$ is a mechanism if $f(v,x,b)$ for all $v\in\Vcal,x\in\Dcal,b\in\Bcal$, i.e, a mechanism computes the allocations given the values and demands of the agents.
% \end{definition}

We say a mapping $f:\Vcal\times\Dcal\times\Bcal\rightarrow\mathbb{R}_+^{NM}$ is an allocation mechanism, i.e, a mechanism computes the allocations given the values and demands of the agents.

We consider Nash Social Welfare (NSW) as the welfare objective for the mechanism designer, which balances social welfare and efficiency. Unlike the egalitarian notion that aims to maximize the utility of the least satisfied agent irrespective of the inefficiency and the utilitarian notion of welfare that maximizes efficiency while disregarding the welfare, NSW sits in-between \shortcite{bertsimas2011price} and achieves a good trade-off. 

A mechanism may further encode agent priorities
via a weights vector $w\in\mathbb{R}_+^{N}$, that are determined solely by the supplier before the agents reveal their requests. 
\begin{definition}[Nash Social Welfare]
Given any mechanism $f$ and $v\in\Vcal,x\in\Dcal,b\in\Bcal$, NSW is defined as
\begin{align}
\begin{gathered}
\NSW(f,v,x,b)\triangleq \Pi_{i=1}^{N}u_i(f,v,x)^{w_{i}},\\
\logNSW(f,v,x,b)\triangleq \log \Big(\NSW(f,v,x,b)\Big)=\textstyle\sum_{i=1}^{N}w_i\log(u_i(f,v,x))\;. 
\end{gathered}
\label{def:NSW}
\end{align}
\end{definition}
We next formalize \emph{exploitability} as the maximum gain an agent may obtain by misreporting its utility function.
\begin{definition}[Exploitability]
Under mechanism $f$ and $v\in\Vcal,x\in\Dcal,b\in\Bcal$, we define the exploitability at agent $i$ as its largest possible utility increase due to misreporting, given that the every other agent reports its true valuation and demand, i.e.,
%\vspace{-5pt}
\begin{align} %\label{def:exploitability}
\exploitability_i(f,v,x,b) \triangleq \max_{v_i',x_i'}u_i\big(f((v_i',v_{-i}),(x_i',x_{-i}),b),v,x\big)-u_i\big(f(v,x,b),v,x\big). \nonumber
\end{align}
\end{definition}
A mechanism $f$ is \textbf{\textit{incentive compatible}} (IC) if~$\forall~i,v,x,b$  $\exploitability_i(f,v,x,b)=0$.

\noindent The goal in incentive compatible mechanism design to promote social welfare while incurring zero or low exploitability for all agents. It is well-known that no mechanism can simultaneously achieve the optimal NSW and perfect incentive compatibility \shortcite{cole2013mechanism}. In this work, we aim to achieve a desirable trade-off among the two metrics. 

% \subsection{Objective}
% The main goal of the mechanism is to maximize fairness \& efficiency of the resource
% allocation that is incentive compatible in the payments-free setting. Since the three objectives cannot be achieved simultaneously (see \shortciteN{cole2013mechanism}), we consider learning approximate mechanisms that achieve a desirable trade-off. Specifically, given $v\in\Vcal,x\in\Dcal,b\in\Bcal$ and an exploitability tolerance $\epsilon\geq 0$:
% \begin{align}\label{eq:obj_instancewise}
%     \max_{f} \quad&\logNSW(f,v,x,b)\\
%     \text{s.t.}\quad&\exploitability_i(f,v,x,b)\leq\epsilon,\quad\forall i=1,\cdots,N.\nonumber
% \end{align}

% \begin{figure}
%   \centering
%   \includegraphics[width=.4\linewidth]{Figures/exploitability_2x2.png}
%   \caption{The utility of agent 1 as it varies the reported preference ratio $v_{1,2}/v_{1,1}$ from 0.1 to 3 (the true ratio is 0.5). Blue line indicates the utility of agent 1 under truthful report, and red line indicates the maximum achievable utility under misreport. Exploitability of PF is shown as their gap. Agent $1$ increases its utility by under-reporting $v_{1,2}/v_{1,1}$.}
%   \vspace{-10pt}
%   \label{fig:exploitability_2x2}
% \end{figure}

\begin{figure}
\centering
\begin{minipage}{.45\textwidth}
  \centering
  \includegraphics[width=0.9\linewidth]{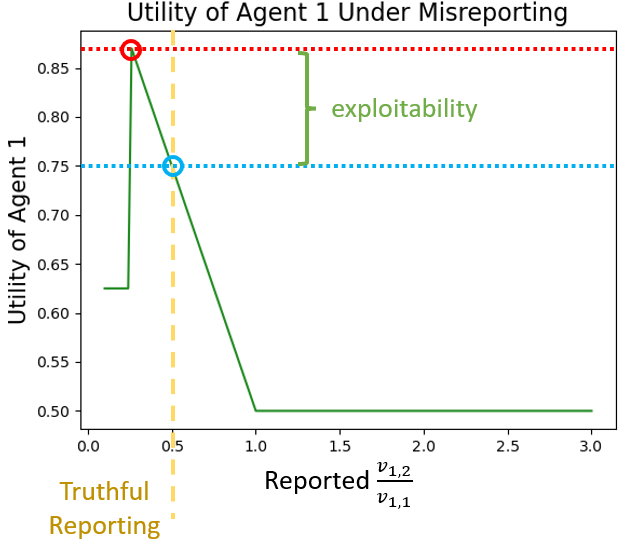}
  \caption{The utility of agent 1 as it varies the reported preference ratio $v_{1,2}/v_{1,1}$ from 0.1 to 3. Blue line indicates the utility of agent 1 under truthful report, and red line indicates the maximum achievable utility under misreport. Exploitability of PF is shown as their gap. Agent $1$ increases its utility by under-reporting $v_{1,2}/v_{1,1}$.}
  \label{fig:exploitability_2x2}
\end{minipage}
\begin{minipage}{.05\textwidth}
\hspace{0pt}
\end{minipage}
\begin{minipage}{.46\textwidth}
  \includegraphics[width=.8\linewidth]{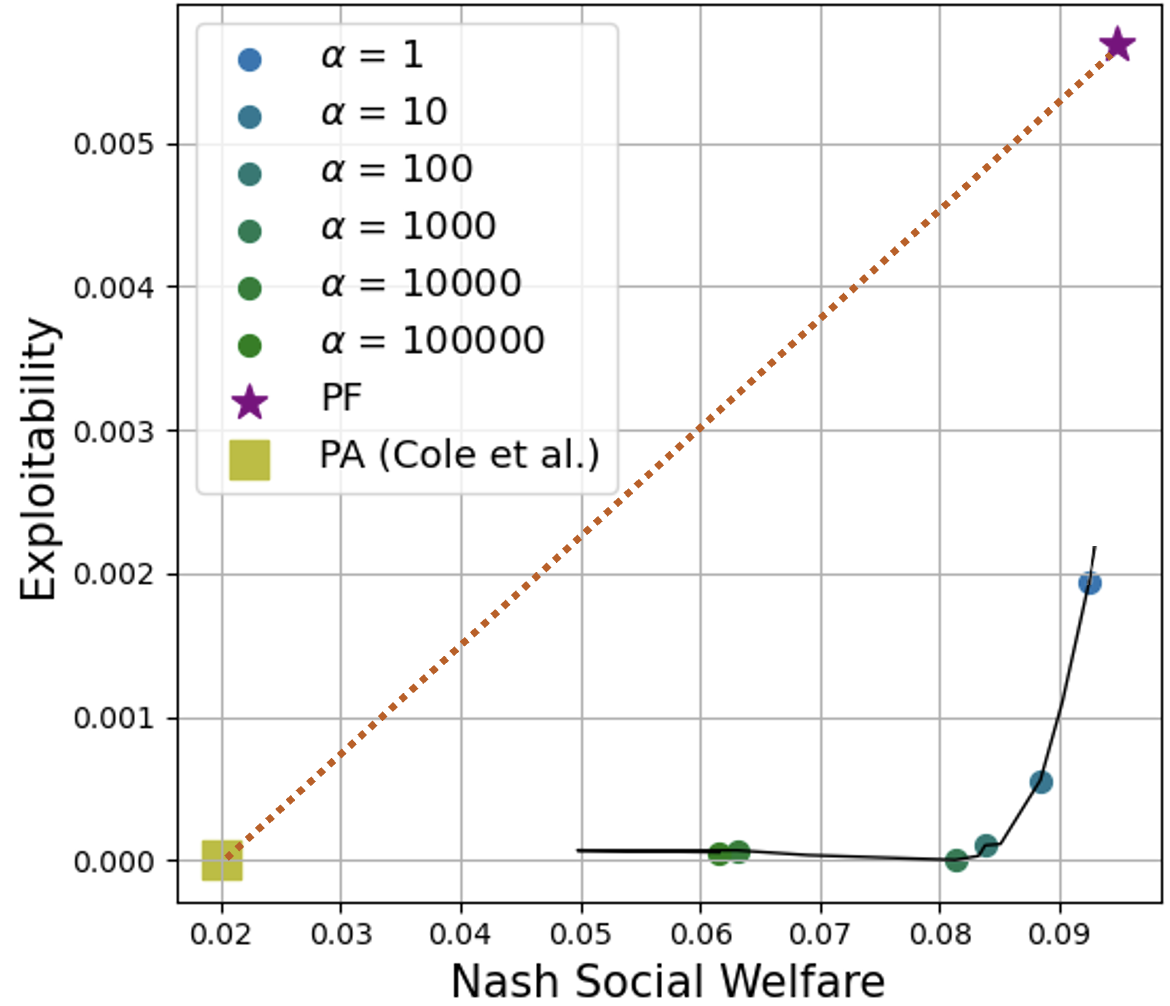}
  \vspace{10pt}
  \caption{Performance trade-off frontier. The solid curve obtained by proposed mechanism \texttt{ExS-Net} over a range of exploitability tolerance (see Definition~\ref{def:epsilon_ic}). The figure illustrates the superior exploitability/NSW trade-off achieved by \texttt{ExS-Net} over the mixture of PA and PF indicated by the dotted line (details in Section \ref{sec:frontier}).
 }
 \label{fig:frontier}
\end{minipage}
\end{figure}

\subsection{State-of-the-art Mechanisms}
\begin{enumerate}
    \item \textit{Proportional Fairness mechanism} ($f^{PF}$): This mechanism was first introduced in \shortciteN{Key97}. Assuming that the agents \textit{truthfully} report their preferences, $f^{PF}$ achieves the optimal NSW~\shortcite{bertsimas2011price}, but the constraint on exploitability is violated even in the simplest instances. 
    % For example, suppose agent-2 reports truthfully, With the dashed line indicating the utility of agent-1 under truthful reporting,  under-reporting $v_{1,2}/v_{1,1}$ increases agent-1's utility up to 17\%.
    As an example, we show how utility gain may arise from misreporting under the PF mechanism in a two-agent two-resource system. When agent 2 reports truthfully, Figure~\ref{fig:exploitability_2x2} plots the utility of agent 1 [cf. \eqref{def:utility}] as it varies the reported preference ratio $v_{1,2}/v_{1,1}$ from 0.1 to 3 (the true ratio is 0.5). With the dashed line indicating the utility of agent 1 under truthful reporting,  under-reporting $v_{1,2}/v_{1,1}$ increases agent 1's utility up to 17\%, which can be a huge incentive in practical applications.
\begin{mechanism*}{Proportional Fairness}
\vspace{-15pt} 
\begin{align*}
f^{PF}(v,x,b)=& \argmin_{a\in\mathbb{R}^{NM}}-{\textstyle\sum}_{i=1}^{N}w_i\log(a_i^{\top}v_i)\label{def:PF}\\
& \hspace{-10pt} \text{s.t.} \ 0\leq a\leq x,\ {\textstyle\sum}_{i=1}^{N}a_{i,m}\leq b_m,\,\,\!\!\forall m.\notag
\end{align*}
\vspace{-25pt} 
\end{mechanism*}

% \item \textit{Partial Allocation mechanism} ($f^{PA}$): Proposed in~\shortciteN{cole2013mechanism}, this mechanism achieves \textit{zero} exploitability by strategically withholding resources. Each agent can only be guaranteed to receive a $1/e$ fraction of the resources that it would receive under the PF mechanism, meaning that there is resource waste and a significant reduction in NSW.

\item \textit{Partial Allocation mechanism} ($f^{PA}$): Proposed in~\shortciteN{cole2013mechanism}, this mechanism is built upon the PF mechanism and withholds resources according to an externality ratio. 
The PA mechanism achieves \textit{zero} exploitability, but each agent can only be guaranteed to receive a $1/e$ fraction of the resources that it would receive under the PF mechanism, meaning that there is resource waste and a significant reduction in NSW. Let $x^{-i}\in\mathbb{R}^{N\times M}$ denote a modified demand matrix such that $x^{-i}_{j}=x_j\in\mathbb{R}^M$ for all $j\neq i$ and $x^{-i}_{i}=0$. We summarize the steps of the PA mechanism as follows.

\begin{mechanism*}{Partial Allocation}
\vspace{-5pt} 
(1) For each agent $i$, calculate $a^{-i,\star}$, the PF allocation outcome that would arise in the absence of agent $i$
\vspace{-2pt} 
\[a^{-i,\star}\triangleq f^{PF}(v,x^{-i},b).\]

\vspace{-5pt} 

(2) Compute the externality ratio $r_i\in\mathbb{R}_+$ for agent $i$ as
\vspace{-7pt} 
\[r_i=\left(\frac{\prod_{j\neq i}u_j(f^{PF}(v,x,b),v,x)^{w_j}}{\prod_{j\neq i}u_j(a^{-i,\star},v,x)^{w_j}}\right)^{1/w_i}.\]
\vspace{-10pt} 

(3) Allocate $f_i^{PA}(v,x,b)\in\mathbb{R}^M$ to agent $i$ according to
\vspace{-5pt} 
\[f_i^{PA}(v,x,b)=r_i f_i^{PF}(v,x,b).\]
\end{mechanism*}

\end{enumerate}
No mechanisms in the literature balance these criteria in the payment-free setting when agents may misreport: PA mechanism is sub-optimal in NSW, whereas PF is optimal in NSW but incurs high exploitability. This dichotomy does not exist in the \textit{auction setting} as payments are used as an enforcement tool for truthful reporting. Classic auction mechanisms including VCG mechanism and Myerson mechanism balance IC with revenue maximization~\shortcite{nisan2007introduction}.

The key challenge that explains the gap in the literature is the difficulty to hand-design mechanisms that trade-off conflicting objectives -- it is easier to design those that satisfy some of them \textit{exactly}. This motivates the work in this paper: We propose a learning based approach to mechanism design that achieves a desirable trade-off between the objectives, in the payment-free setting. As illustrated in Figure~\ref{fig:frontier}, by adjusting a tunable weight parameter, we learn a mechanism (\texttt{ExS-Net}) that achieves a NSW-exploitability trade-off frontier that significantly improves over PF and PA. Simulation details will be discussed later in Section~\ref{sec:frontier}.

%\vspace{-10pt}
\section{Learning Payment-free Mechanism}\label{sec:mechanisms}
In this section, we design a novel ``state-augmentation with an artificial agent'' based neural architecture that simulates ``money-burning''~\shortcite{HR08} in the ``hardware'', i.e., intentionally withholding of resources as an implicit form of payment. %We define the \texttt{ExS-Net} architecture in Section~\ref{sec:architecture}.  %We then present the loss and algorithm used for training $\omega$ in Section~\ref{sec:training} and the procedure of using \texttt{ExS-Net} for inference in Section~\ref{sec:inference}.

\subsection{\texttt{ExS-Net} -- Strategic Resource Withholding}\label{sec:architecture}
\begin{figure}
  \centering
  \includegraphics[width=.8\linewidth]{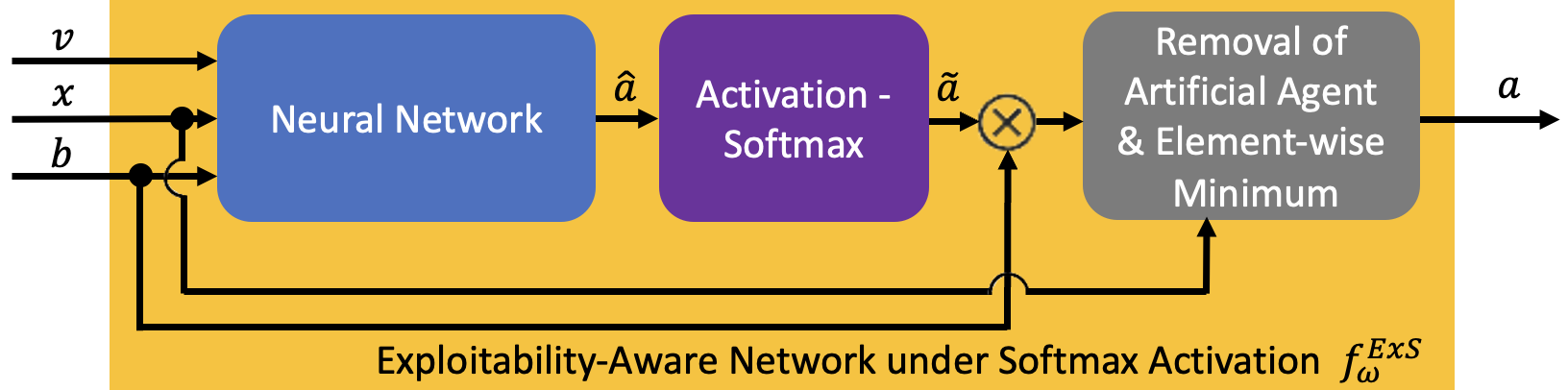}
  \caption{Exs-Net Pipeline}
  \vspace{-10pt}
  \label{fig:flow}
\end{figure}

We name the mechanism \texttt{ExS-Net}, which stands for \textbf{Ex}ploitability-Aware Network with \textbf{S}oftmax activation, and provide a schematic representation in Figure~\ref{fig:flow}. The \texttt{ExS-Net} mechanism can be represented as a function $f^{\omega}:\Vcal\times\Dcal\times\Bcal\rightarrow\mathbb{R}^{NM}_+$ parameterized by neural network parameter $\omega$. Specifically, we employ a neural network that maps $v,x,b$ to an initial allocation weight vector $\hat{a}\in\mathbb{R}^{(N+1)M}$ (rather than $\mathbb{R}^{NM}$). As pointed out in \shortciteN{HR08}, IC cannot be achieved in the payment-free setting without resource withholding (``money burning''). Motivated by this knowledge, we expand the space of allocations to introduce a synthetic agent that receives the proportion of the allocation that will eventually be withheld. This synthetic agent enlarges the allocation dimension in $\hat{a}$. 
% Instead of directly penalizing the exploitability gap associated with misreporting, ExS-Net uses parameter augmentation to dilute its effects in the form of a softmax function applied to $\hat{a}$ along the direction of \emph{true} agents 
Through the use of a softmax function applied to $\hat{a}$ along the direction of agents, ExS-Net calculates the allocation ratio 
$\tilde{a}\in\mathbb{R}^{(N+1)M}$ such that $\sum_{i=1}^{N+1}\tilde{a}_{i,m}=1$, i.e., 
%
%\vspace{-5pt}
\begin{equation*}
% \label{eq:ExS-Net}
\tilde{a}_{i,m}=\exp(\hat{a}_{i,m})/\big({\textstyle\sum}_{i'=1}^{N+1}\exp(\hat{a}_{i',m})\big)\text{ for all }m.
\end{equation*}
As the fraction of resource $m$ assigned to agent $i$, the quantity $\tilde{a}_{i,m}$ leads to  allocation $a_{i,m}=\min\{\tilde{a}_{i,m}b_{m},x_{i,m}\}$. We note that our design is modular in that any neural network (feed-forward, CNN, Transformers) can be used as the hidden layer and the money-burning hardware only modifies the output/activation layer. 

\subsection{Training \texttt{ExS-Net} Mechanism}\label{sec:training}

Assuming that the values and demands of the agents and the budgets on the resources follow a joint distribution $F$, we use samples from $F$ to train a mechanism that optimize NSW while staying approximately incentive compatible in expectation. To this end, we define the notion of $\epsilon$-incentive compatibility in expectation.

\begin{definition}\label{def:epsilon_ic}[$\epsilon$-Incentive Compatibility]
A mechanism $f$ is $\epsilon$-incentive compatible over distribution $F$ if
\begin{align}
    \mathbb{E}_{(v,x,b)\sim F}[\exploitability_i(f,v,x,b)] \leq \epsilon,\quad \forall i.\label{eq:epsilon_ic}
\end{align}
\end{definition}

\begin{algorithm}[h]
\SetAlgoLined
%\color{red}
\textbf{Input:} Initial network parameter $\omega^{[0]}$, dual variables $\{\gamma_i^{[0]}\}_{i=1}^{N}$, training dataset $\{(v^l,x^l,b^l)\}_{l=1}^{L}$, batch size $s$, training iterations $K$, primal and dual learning rate $\alpha,\beta$

\textbf{Output:} Network parameter $\omega^{[K]}$.

 \For{$k=0,1,\cdots,K-1$}{
  
    1) Randomly draw sample index set $\Scal^{[k]}$ with $|\Scal^{[k]}|=s$ and compute empirical logNSW and exploitability
    \vspace{-5pt}
    \begin{align*}
        \widehat{\logNSW}^{[k]} = \textstyle{\sum}_{l\in\Scal^{[k]}}\logNSW(f^{\omega^{[k]}},v^l,x^l,b^l),\\
        \widehat{\exploitability}_i^{[k]} = \textstyle{\sum}_{l\in\Scal^{[k]}}\exploitability_i(f^{\omega^{[k]}},v^l,x^l,b^l),\quad\forall i\\
    \end{align*}
    \vspace{-30pt}

    2) Neural network parameter update: 
    \vspace{-5pt}
    \begin{align*}
        \omega^{[k+1]}=\omega^{[k]}-\alpha \nabla_{\omega^{[k]}}\big(\sum_{i}\hspace{-1pt}\gamma_i^{[k]}\widehat{\exploitability}_i^{[k]}-\widehat{\logNSW}^{[k]}\big)
    \end{align*}

    \vspace{-10pt}
    3) Dual variable update:
    \vspace{-5pt}
    \begin{align*}
        \gamma_{i}^{[k+1]}=\Pi_{+}\Big(\gamma_{i}^{[k]}+\beta\big(\widehat{\exploitability}_i^{[k]}-\epsilon\big)\Big),\quad\forall i
    \end{align*}
    \vspace{-10pt}
 }
\caption{Training ExS-Net}
\label{Alg:training}
\end{algorithm}

We define learning $\omega$ as the maximization of expected NSW with an $\epsilon$-incentive compatibility constraint. The learning objective is given as follows.
\begin{tcolorbox} [colback=red!5!white,colframe=white!25!black,title=Learning Objective]
\vspace{-10pt}
\begin{align}\label{eq:ExPF-Net_train_obj}
\max_{\omega}\,\,&\mathbb{E}_{(v,x,b)\sim F}[\logNSW(f^{\omega},v,x,b)]\quad\text{s.t.}\,\, \mathbb{E}_{(v,x,b)\sim F}[\exploitability_i(f^{\omega},v,x,b)]\leq\epsilon,\,\forall i.
\end{align} 
\end{tcolorbox}
\noindent In practice, one may not have access to the distribution $F$, but instead a training set of values, demand, and budgets $\{(v^l,x^l,b^l)\sim F\}_{l=1}^L$.
We train the mechanisms with the finite dataset via empirical risk minimization (ERM) by forming the sample-averaged estimates of the expected NSW and exploitability.
\begin{align}
\textstyle{\max}_{\omega}\,\,{\textstyle\sum}_{l=1}^{L}\logNSW(f^{\omega},v^l,x^l,b^l)\quad\text{s.t.}\,\, {\textstyle\sum}_{l=1}^{L}\exploitability_i(f^{\omega},v^l,x^l,b^l)\leq\epsilon,\,\forall i.
\label{eq:ExPF_obj_ERM}
\end{align}
We note that \eqref{eq:ExPF_obj_ERM} deviates from the conventional supervised learning objectives -- we do not need paired ground truth data. Instead, the objective \eqref{eq:ExPF_obj_ERM} can be evaluated and optimized with a dataset on truthfully reported valuations and demands only, which can be elicited under an existing incentive compatible mechanism such as the PA mechanism.

We optimize \eqref{eq:ExPF_obj_ERM} with respect to $\omega$ by using a simple primal-dual gradient descent-ascent algorithm to find the saddle point of the Lagrangian. We present the training scheme in Alg.~\ref{Alg:training}, where $\Pi_+$ denotes the projection of a scalar to the non-negative range. The dual variable $\gamma_i\in\mathbb{R}_+$ is associated with the $i_{\text{th}}$ exploitability constraint in \eqref{eq:ExPF_obj_ERM}. In essense, the neural network parameter is updated to optimize the sum of log NSW and exploitability, with the weight of exploitability adjusted dynamically according to the level of constraint violation. 

\subsection{Inference using \texttt{ExS-Net}}\label{sec:inference}

We use $\omega^{\star}$ to denote the neural network parameter returned in the training procedure.
In the inference phase, we need to generate allocations given budgets and agent-reported demands \& valuations using \texttt{ExS-Net} parameterized by $\omega^{\star}$.

\begin{mechanism*}{\texttt{ExS-Net}}
(1) Given input budget $b$, reported valuation $v$, and reported demands $x$, compute $\hat{a}\in\mathbb{R}^{(N+1)M}$ as output of neural network parameterized by $\omega^{\star}$\\
(2) Soft-max function
\begin{align*}
    \tilde{a}_{i,m} = \exp(\hat{a}_{i,m})/\big({\textstyle\sum}_{i'=1}^{N+1}\exp(\hat{a}_{i',m})\big)\text{ for all }m.
\end{align*}
3) Determine allocation $f^{\omega^{\star}}(v,x,b)=a \in\mathbb{R}^{NM}$ such that
\[a_{i,m}=\min\{\tilde{a}_{i,m}b_m, x_{i,m}\}\text{ for all }i=1,\cdots,N, m=1,\cdots,M.\]
Discard remaining resources.
\end{mechanism*}

We note that while \texttt{ExS-Net} is given the ability to discard resources with the aim of achieving approximate incentive compatibility, we observe through numerical simulations that the quantity it learns to discard is small. This is reflected by the near-optimal efficiency and high NSW of the mechanism showcased in Section~\ref{sec:simulation}. In other words, we can reduce the exploitability (in the experiments over~$80\%$ reduction is observed on average) at the cost of a small compromise in efficiency and NSW.

\section{Generalization Bounds}\label{sec:generalization_bounds}
We establish the convergence of the learned mechanisms by bounding the sub-sampling error by a sub-linear function of the batch size, which matches recent rates for auctions \shortcite{dütting2022optimal} and ensures that the objective of \eqref{eq:ExPF_obj_ERM} converges to \eqref{eq:ExPF-Net_train_obj} with sufficient samples. 

\noindent For mechanism $f$, we define the generalization errors with $L$ samples as
\vspace{-5pt}
\begin{align*}
    \varepsilon_{\logNSW}(f,L)&=-\mathbb{E}_{(v,x,b)_\sim F}[\logNSW(f,v,x,b)] +{\textstyle\sum}_{l=1}^L\logNSW(f,v^l,x^l,b^l),\notag\\
    \varepsilon_{\mathbf{exp},i}(f,L)&=\mathbb{E}_{(v,x,b)_\sim F}[\exploitability_i(f,v,x,b)] -{\textstyle\sum}_{l=1}^L\exploitability_i(f,v^l,x^l,b^l).
\end{align*}

\begin{thm}[Generalization Bound]\label{thm:GB}
Consider \texttt{ExS-Net} $f^{\omega}$ mechanisms parameterized by a neural network with~$R$ hidden layers,~$K$ nodes per hidden layer, ReLU activation, a total of $d$ parameters, and the vector of all model parameters $\| \omega\|_1 \leq \Omega$. With probability at least~$1-\delta$, 
\begin{align*}
\max\{\varepsilon_{\logNSW}(f^{\omega},L),\varepsilon_{\mathbf{exp},i}(f^{\omega},L)\}\leq \Ocal\Big(\!\frac{\sqrt{Rd \log(\! LN\Omega \max\{K,MN\})}}{L}\!+N\sqrt{\frac{\log(1/\delta)}{L}}\,\Big)\hspace{-1pt}.
\end{align*}
\end{thm}

\vspace{-5pt}
The result provides that the generalization error decays at rate $\Ocal(L^{-1/2})$ where $L$ is the training sample size.
In auction design with payment, RegretNet, the state-of-the-art learned mechanism proposed in \shortciteN{dütting2022optimal}, also achieves a $\Ocal(L^{-1/2})$ rate, which we  match in the non-payment setting. The proof of Theorem~\ref{thm:GB} is inspired by \shortciteN{dütting2022optimal}, but extended to handle: i) the NSW objective in the absence of payment; ii) the special activation functions proposed in \texttt{ExS-Net}.

\section{Experiments}\label{sec:simulation}
% \textcolor{blue}{NEED to work on this section onwards}
In this section, we demonstrate the effectiveness and robustness of the proposed \texttt{ExS-Net} mechanism from various perspectives. First, we evaluate the social welfare, efficiency, and exploitability of the mechanism in systems with different numbers of agents and resources. Second, we study distribution mismatch, a situation where the mechanism is required to perform under a test distribution different from that observed during training. Finally, we adjust the parameter $\epsilon$ in \eqref{eq:ExPF_obj_ERM}, which controls the trade-off between social welfare and exploitability, and show the superior trade-off frontier. Additional experiments on problems of varying scales and budget levels omitted from this version due to space constraints are available in the appendix of the longer version~\shortcite{zeng2023near}.

\subsection{Simulation Setup}
% \textcolor{blue}{Describe the setup}

We use a four-layer neural network as the function approximation for \texttt{ExS-Net} in this work. The specific simulation setup including data distribution, baselines, and evaluation metrics are discussed below.

\noindent\textbf{Data Generation.}
In all experiments, values and demands follow uniform and Bernoulli uniform distributions, respectively, within the range $[0.1, 1]$. Specifically, we sample
\begin{gather}
v_{i,m}\sim\text{Unif}(0.1,1)\;,\widetilde{x}_{i,m}\!\sim\!\text{Unif}(0.1,1\!),\quad\widehat{x}_{i,m}\!\sim\!\text{Bern}(0.5)\;,   x_{i,m}\!\!=\!\widetilde{x}_{i,m}\widehat{x}_{i,m}.\label{eq:data_generation}
\end{gather}

We make $x_{i,m}$ a product of Bernoulli and uniform random variables as we would like to represent the case where not all agents request all resources.
Unless noted otherwise, the budget for every resource is set to $\frac{N}{2}$, for number of agents $N$. This creates competition for resources in expectation. The agent weights are set to 1. Test data is sampled from the same distributions as training data except in Section~\ref{sec:exp:dist_mismatch}.   

\vspace{.1cm}
\noindent\textbf{Baselines.}
We evaluate against the PF \& PA mechanisms. Both PF and PA are hand-designed mechanisms that do not require training, but are time-consuming during inference since they solve convex optimization programs. We use an interior-point solver in this work, which is guaranteed to converge within polynomial time, i.e. the time to obtain a solution up to precision $\varepsilon$ is no more than some polynomial function of $\varepsilon$, $N$, and $M$. However, the exact complexity is unknown, which we should expect to be worse than $\Omega((NM)^3)$ \shortcite{renegar1988polynomial} (as $\Ocal((NM)^3)$ is the time it would take for the interior-point method to converge if the program were a linear program).
We show in Table~\ref{table:time} the training and inference time of the proposed mechanism compared with PF and PA in an actual ten-agent three-resource simulation. While requiring computation in the training process, \texttt{ExS-Net} produces allocations much faster during inference.

A mixture of PA and PF provides a stronger benchmark in terms of trading off both NSW and exploitability. Given $\rho\in[0,1]$, we consider the mechanism $f^{\text{mixture}}$ below. Varying $\rho$ between $[0,1]$ interpolates between PF and PA in expectation. We set $\rho=1/2$ for comparison in Section~\ref{sec:exp:scalingsystem}.
\begin{align*}
\!\!r\hspace{-2pt}\sim\hspace{-2pt}\text{Bern}(\rho),\, f^{\text{mixture}}(v,x,b)\!=\! \begin{cases} \!
f^{PF}(v,x,b), & \!\!\text{if }r\!=\!1 \! \\
\!f^{PA}(v,x,b), & \!\!\text{if }r\!=\!0
\end{cases}
% \label{def:mixture_PFPA}
\end{align*}

\vspace{-5pt}
\begin{table}[!h]
\centering
\setlength{\tabcolsep}{5pt}
\begin{tabular}{ccccc}
\toprule
\makecell{Mechanism}  &  \makecell{Training Time \\(Theory)} & \makecell{Inference Time \\(Theory)} &  \makecell{Training Time \\(Simulation)} & \makecell{Inference Time \\(Simulation)} \\
        \midrule
        Proportional Fairness & 0 & Worse than $\Omega((MN)^3)$ & 0 & 40.1 \\
        Partial Allocation & 0 & Worse than $\Omega((MN)^3)$ & 0 & 310.0 \\
        \texttt{ExS-Net} (Proposed) & $\Ocal(MN)$ & $\Ocal(MN)$ & \textbf{1} & \textbf{1} \\
        \bottomrule
        \bottomrule
        \end{tabular}
\caption{Training and inference time in a ten-agent three-resource allocation problem (normalized).}
\label{table:time}
\end{table}

\noindent\textbf{Evaluation Metrics.}
Mechanisms are evaluated on NSW \eqref{def:NSW}, exploitability, and efficiency
\[\exploitability(f,v,x,b) =(1/N)\textstyle\sum_{i=1}^{N}\exploitability_i(f,v,x,b),\quad\efficiency(f,v,x,b) = \frac{\sum_{m=1}^{M}\sum_{i=1}^{N}f_{i,m}(v,x,b)}{\sum_{m=1}^{M}b_m}.\]

% \[\efficiency(f,v,x,b\!) \!=\!\! \left({\sum_{m=1}^{M}\!\sum_{i=1}^{N}\!f_{i,m}(v,x,b)}\!\!\right)\!\!/ \!\left({\sum_{m=1}^{M}\!b_m}\!\!\right)\!.\]

It is desirable for mechanisms to have a high efficiency to reduce the waste of resource, although efficiency merely is a byproduct of NSW in the \texttt{ExS-Net} training objective. Since budget may exceed total demand, maximum  efficiency can be smaller than 1. The PF mechanism is fully efficient (i.e. all available resources are used for sufficient demand), and hence serves as a suitable reference point.

\subsection{Scaling System Parameters}\label{sec:exp:scalingsystem}
\begin{figure*}[!ht]
  \centering
  \includegraphics[width=\linewidth]{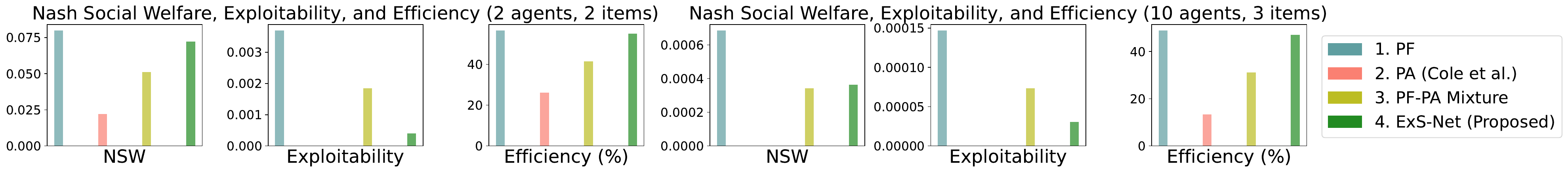}
  
  \caption{Mechanism performance in 2x2 and 10x3 systems}
  \vspace{-7pt}
  \label{fig:2x2_10x3}
\end{figure*}

Figure~\ref{fig:2x2_10x3} shows the mechanism performance on two-agent two-resource and ten-agent three-resource allocation problems, with the tolerance on exploitability set with respect to that of the PF mechanism to $10^{-3}$ and $10^{-4}$, respectively. 
The 2x2 system is the smallest non-trivial case, while the 10x3 system is the largest considered in the recent works \shortcite{dütting2022optimal,ivanov2022optimal} on auction design with payment.\looseness=-1

The figure shows that \texttt{ExS-Net} achieves an advantageous trade-off between PF and PA: it consistently reduces the exploitability of PF by over at least 80\% while achieving remarkable NSW and almost full efficiency. Compared with PA, \texttt{ExS-Net} significantly improves the efficiency and NSW. In addition, \texttt{ExS-Net} outperforms the interpolated mixture of PF and PA under all three metrics.

\begin{figure*}[!ht]
  \centering
  \includegraphics[width=\linewidth]{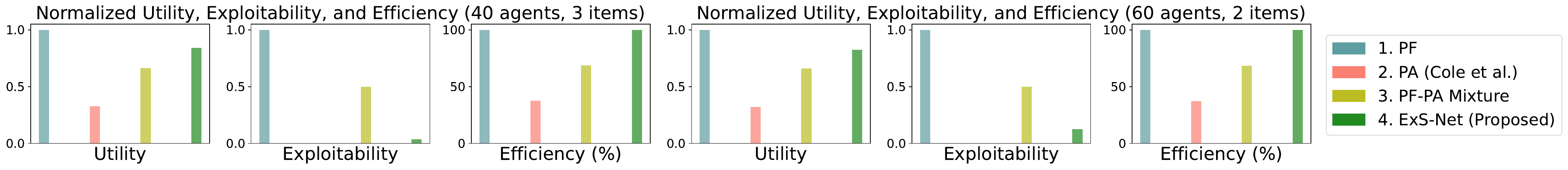}
  \caption{Mechanism performance in 40x3 and 60x2 systems}
  \vspace{-10pt}
  \label{fig:largersimulations}
\end{figure*}

In Figure.~\ref{fig:largersimulations}, we conduct experiments in larger systems with 40-60 agents and plot metrics including the averaged per-agent utility, exploitability, and efficiency. The improvement is consistent -- each agent on average gets over 82\% of the utility it would get under the PF mechanism while exploitability is reduced by 88-96\%.\looseness=-1

\subsection{Distribution Mismatch}\label{sec:exp:dist_mismatch}

We investigate performance when the distribution of values and demands the mechanism is trained differs from that on which it is tested. This mismatch may occur due to measurement error or more pernicious sources such as strategically misrepresenting preferences. While we may limit such behavior by running an IC mechanism (such as the PA mechanism) to collect training samples, in this section we numerically examine the robustness of the mechanisms from Section \ref{sec:mechanisms} if untruthful training samples are present.

Recall that $F$ denotes the true joint distribution and $F’$ denotes the distribution of the training samples. Further denote by $\omega^{\star}(F')$ the optimal solution to the ERM problem \eqref{eq:ExPF_obj_ERM} under samples drawn from distribution $F'$. With a large discrepancy between $F$ and $F$, the performance of $\omega^{\star}(F')$ under the true distribution $F$, measured by 
$\mathbb{E}_{ F}[\NSW(\omega^{\star}(F'),v,x,b)]$ and $\mathbb{E}_{F}[\exploitability(\omega^{\star}(F'),v,x,b)]$, can theoretically be worse than that of $\omega^{\star}(F)$.
Nonetheless, empirically we observe robustness: $\mathbb{E}_{F}[\NSW(\omega^{\star}(F'),v,x,b)]$ and $\mathbb{E}_{F}[\exploitability(\omega^{\star}(F'),v,x,b)]$ closely match $\mathbb{E}_{F}[\NSW(\omega^{\star}(F),v,x,b)]$ and $\mathbb{E}_{F}[\exploitability(\omega^{\star}(F),v,x,b)]$ across two common types of distribution mismatch.

\noindent\textbf{Randomly perturbed training samples}
In a $2\times 2$ system we suppose that the distribution $F$ follows \eqref{eq:data_generation} while the training samples $\{(v^l,x^l)\}_l$ are generated under the perturbation of a heavy-tailed Cauchy random variable, which is often used to model unobserved strategic behavior \shortcite{taleb2010black}. %Details can be found in Appendix~\ref{sec:supp:dist_mismatch}, along with a distribution density plot of the perturbed samples.

\noindent\textbf{Adversarially generated training samples}
Again, we consider a $2 \times 2$ setting where the true distribution $F$ is described in \eqref{eq:data_generation}.
Suppose that the training dataset is composed of historical valuations and demands collected through past interactions with the agents. If the agents believe that allocations are made by the PF mechanism in these past interactions, they may have reported strategically with the aim of maximizing their own utility, which means that the training samples $\{(v^l,x^l)\}_l$ will take the following form:
\begin{align}
    \bar{v}_{i,m}^l \sim \text{Unif}(0.1,1),\quad \bar{x}_{i,m}^l \sim \text{Unif}(0.1,1)\quad v_{i}^l,x_{i}^l\hspace{-2pt}=\hspace{-2pt}\arg\max_{v_i',x_i'}u_i\hspace{-1pt}(f((\hspace{-1pt}v_i',\hspace{-1pt}\bar{v}_{-i}^l),(x_i',\bar{x}_{-i}^l),b),\bar{v}^l,\bar{x}^l).
\label{eq:dist_mismatch_adversarial}
\end{align}

Table~\ref{table:misreport_metrics} shows the performance of the proposed mechanisms trained using \eqref{eq:dist_mismatch_adversarial}, which closely tracks that under truthful training data across all evaluation metrics. 
This experiment demonstrates the robustness of the learning-based mechanisms to mismatch between training and inference distribution.
% Thus, the learning-based mechanisms are robust to mismatch between training and inference distribution, 
% and corroborates the theoretical results in Section \ref{sec:distribution_shifts}.

\begin{table*}[!h]
\centering
\setlength{\tabcolsep}{5pt}
\begin{tabular}{cccc}
\toprule
\makecell{Mechanism}  &  \makecell{NSW} & \makecell{Exploitability} & \makecell{Efficiency (\%)} \\
        \midrule
        Proportional Fairness Mechanism & 8.00e-2$\pm$1.5e-2 & 3.70e-3$\pm$1.2e-3 & 56.6$\pm$3.9 \\
        Partial Allocation Mechanism & 2.22e-2$\pm$2.1e-3 & 0$\pm$0 & 26.2$\pm$2.5 \\
        PF-PA Mixture & 5.11e-2$\pm$6.6e-3 & 1.85e-3$\pm$6.2e-4 & 41.4$\pm$3.6 \\
        \texttt{ExS-Net} (Trained on Truthful Data)  & 7.24e-2$\pm$4.3e-3 & 4.04e-4$\pm$9.6e-5 & 55.0$\pm$4.2 \\
        \texttt{ExS-Net} (Trained on Randomly Perturbed Data)  & 7.26e-2$\pm$1.0e-2 & 4.78e-4$\pm$2.9e-4 & 55.2$\pm$1.4 \\
        \texttt{ExS-Net} (Trained on Adversarial Data)  & 7.01e-2$\pm$2.0e-2 & 5.25e-4$\pm$1.7e-4 & 54.9$\pm$2.6 \\
        \bottomrule
        \bottomrule
        \end{tabular}
\vspace{-2pt}
\caption{Performance of mechanism trained under data containing untruthful reporting of agents' preferences.}
\vspace{-5pt}
\label{table:misreport_metrics}
\end{table*}

\subsection{Trade-Off Frontier between Nash Social Welfare and Exploitability}\label{sec:frontier}

Instead of training the mechanisms on the constrained objective \eqref{eq:ExPF-Net_train_obj}, we can alternatively consider the unconstrained program below, which allows us to directly control the weight of exploitability relative to NSW.\looseness=-1
\begin{align}
\max_{\omega}\,\,&\mathbb{E}_{(v,x,b)\sim F}[\logNSW(f^{\omega},v,x,b)+\alpha{\textstyle\sum}_{i=1}^N\exploitability_i(f^{\omega},v,x,b)].\label{eq:training_obj_unconstrained}
\end{align}
In Figure~\ref{fig:frontier}, we plot the performance frontier of the \texttt{ExS-Net} mechanism under as $\alpha$ varies across orders of magnitude in a two-agent two-resource system. 
% The figure shows that in terms of the trade-off between exploitability and NSW, the proposed ExS-Net outperforms the mixture of PA and PF mechanisms by a significant margin.
The figure shows that the probabilistic mixture of PF and PA, indicated by the dotted line, falls inside the frontier by a large margin, indicating that the proposed \texttt{ExS-Net} significantly outperforms the mixture of PA and PF mechanisms in terms of the trade-off between exploitability and NSW.

% \begin{figure}[!ht]
%  \centering
%  \includegraphics[width=.5\linewidth]{Figures/frontier.png}
%  \caption{Performance trade-off. The solid curve obtained by ExS-Net clearly has a better trade-off than the mixture of PA and PF indicated by the dotted line.
%  }
%  \label{fig:frontier}
% \end{figure}

\section{Conclusion \& Future Directions}
We studied learning of fair resource allocation mechanisms without payments. We designed a modular neural network architecture that can take in general layouts (feed-forward, transformers, CNN etc) in the hidden layers and tack-on the strategic resource withholding in the activation layer to achieve the desired trade-off. This flexibility increases the generality and applicability of the proposed methods for approximately fair and incentive compatible mechanisms in the payment-free setting. The payment free setting required changes in the ``hardware'' and is not a simple flip of the objective in~\shortciteN{dütting2022optimal}. Additionally, the welfare in the objective posed additional challenges in characterizing the generalization bound, adding to the literature on learning mechanisms~\shortcite{dütting2022optimal,ivanov2022optimal,mishra2022eef1}. We conducted an extensive empirical study of the proposed mechanisms and compared with the state-of-the-art baselines that lie at the end of the spectrum. The results positively affirm that the desired trade-offs are achieved by \texttt{ExS-Net}.  

% Future directions include a systematic way to evaluate the exploitability of the PF mechanism. For small systems, it is possible to sweep through the values to evaluate the exploitability, but this method is not scalable. Techniques from discipline parameterized programs~\shortcite{agrawal2019differentiable} are promising in this regard. While \texttt{ExS-Net} achieves a desirable trade-off it is not clear if this is the best possible in the performance frontier. Future directions on further changes in the activation layer to improve the trade-off are promising. 

% \textcolor{blue}{\section{Outline for Appendix}
% \begin{itemize}
%     \item Include the two mechanisms
%     \item Include the proof
% \end{itemize}}

% \clearpage

% \section*{Impact Statement}\label{sec:impact}
% This paper presents work whose goal is to advance the field of Machine Learning. There are many potential societal consequences of our work, none which we feel must be specifically highlighted here.

%\textcolor{blue}{\textbf{Sec:Additional Results}: (i) Comment on how the methods can be adopted to indivisble resource, what are the considerations, modifications, interpretations etc. (ii) Robustness of NN performance results, (iii) Generalization bounds, (iv) Comments on extensions to other utilities, (v) Connections to learning with money } \\

\section*{Disclaimer}
This paper was prepared for informational purposes in part by
the Artificial Intelligence Research group of JP Morgan Chase \& Co and its affiliates (``JP Morgan''),
and is not a product of the Research Department of JP Morgan.
JP Morgan makes no representation and warranty whatsoever and disclaims all liability,
for the completeness, accuracy or reliability of the information contained herein.
This document is not intended as investment research or investment advice, or a recommendation,
offer or solicitation for the purchase or sale of any security, financial instrument, financial product or service,
or to be used in any way for evaluating the merits of participating in any transaction,
and shall not constitute a solicitation under any jurisdiction or to any person,
if such solicitation under such jurisdiction or to such person would be unlawful.

% Please don't exchange the bibliographystyle style
\bibliographystyle{wsc}
% AUTHOR: Include your bib file here
\bibliography{references}

\section*{AUTHOR BIOGRAPHIES}

% {\normalsize Author biographies are mandatory for all authors and given in the same sequence as in the running head. Use a smaller font size (9pt) as it is set up in this template. Giving an email address is mandatory and the author's name is set in bold capitals (see the examples below). Separate the authors by an empty line with the same style. Give only one paragraph per author, but sufficient information to understand the author's position and scientific background.}\\

\noindent {\bf \MakeUppercase{Sihan Zeng}} is a Research Scientist at JPMorgan AI Research with expertise in optimization and reinforcement learning. His email address is  \email{sihan.zeng@jpmchase.com}.\\

\noindent {\bf \MakeUppercase{Sujay Bhatt}} is a Research Lead at JPMorgan AI Research with expertise in bandits, reinforcement learning, and optimization. His email address is  \email{sujay.bhatt@jpmchase.com}.\\

\noindent {\bf \MakeUppercase{Eleonora Kreacic}} is a Research Lead at JPMorgan AI Research, with interest and expertise in generative models and graph theory. Her email address is  \email{eleonora.kreacic@jpmorgan.com}.\\

\noindent {\bf \MakeUppercase{Parisa Hassanzadeh}} is a AI Research Scientist at Samsung SDS. Her research broadly spans telecommunications, machine learning, and the applications of AI in finance. Her email address is  \email{parisah@nyu.edu}.\\

\noindent {\bf \MakeUppercase{Alec Koppel}} is a Research Lead at JPMorgan AI Research with expertise in optimization, machine learning, and sequential decision making. His email address is  \email{alec.koppel@jpmchase.com}.\\

\noindent {\bf \MakeUppercase{Sumitra Ganesh}} is a Research Director at JPMorgan AI Research, leading a team on AI Agents, multi-agent simulations, and sequential decision making. Her email address is  \email{sumitra.ganesh@jpmorgan.com}.\\

\clearpage

\appendix
\onecolumn
\vbox{%
\hrule height 1pt
\hsize\textwidth
\linewidth\hsize
% \hrule height 4pt
\vskip 0.25in
\centering
{
\Large\bf
{Learning Payment-Free Resource Allocation Mechanisms \\ 
Supplementary Materials} \par}
\vskip 0.2in
\hrule height 1pt
\vskip 0.2in
}

% \input{Appendix_RelatedWork}

% \input{Appendix_Example}

% \input{Appendix_PAMechanism}

% \onecolumn

\section{Additional Experimental Results}\label{sec:additional_experiments}

\subsection{Additional Experiments with Various System Parameters}\label{sec:additional_experiments:systemparams}

\begin{figure*}[!ht]
  \centering
  \includegraphics[width=\linewidth]{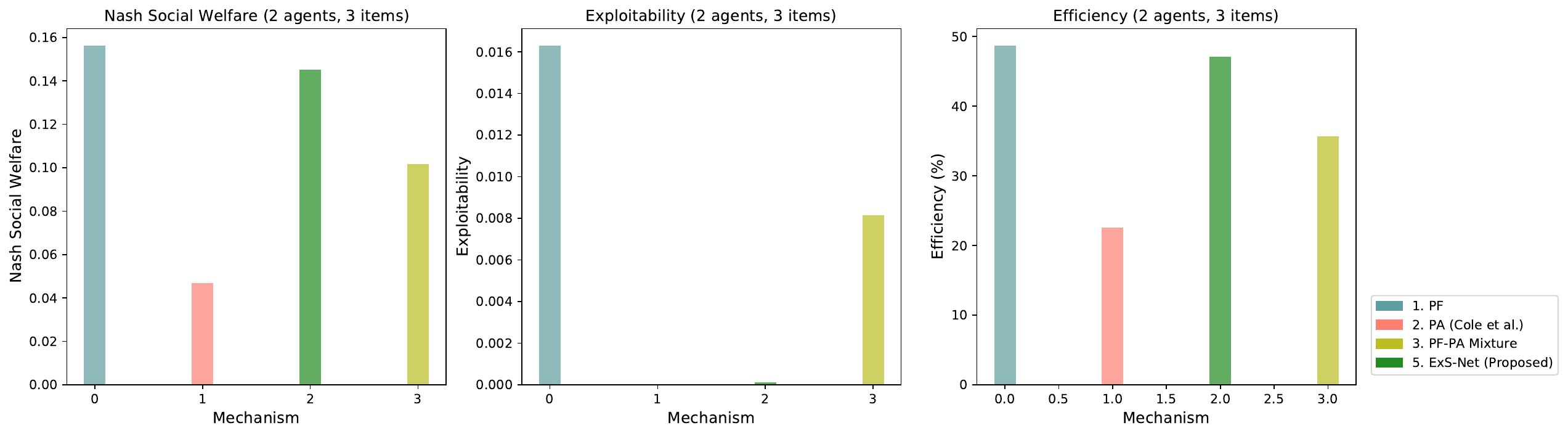}
  \includegraphics[width=\linewidth]{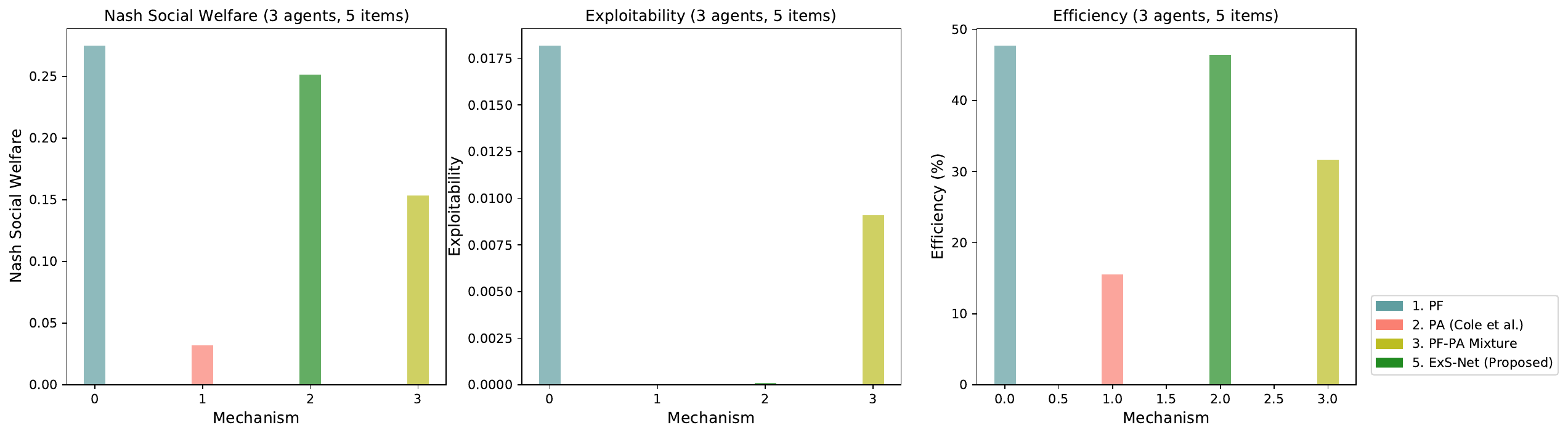}
  \includegraphics[width=\linewidth]{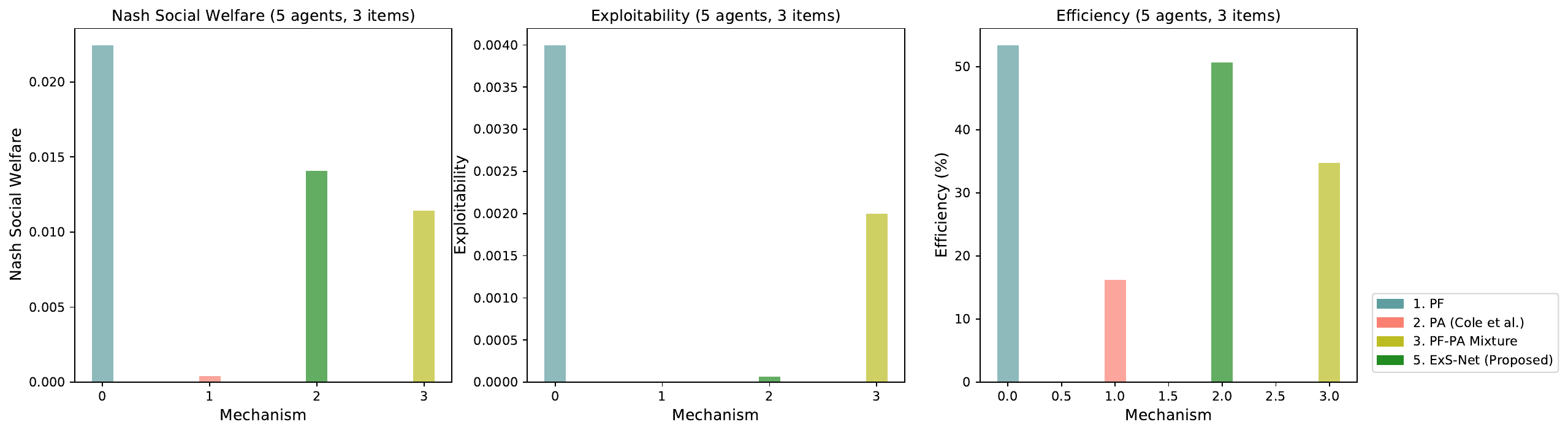}
  \caption{Mechanism performance in 2x3, 3x5, and 5x3 systems}
  \label{fig:moresimulations}
\end{figure*}

We plot the performance of ExS-Net in 2-agent 3-resource, 3-agent 5-resource, and 5-agent 3-resource problems in Figure.~\ref{fig:moresimulations}. 
The figure shows the consistently superior performance of learning-based ExS-Net in achieving a much more desirable trade-off between NSW and exploitability.

\begin{figure}[!ht]
  \centering  
  \includegraphics[width=.8\linewidth]{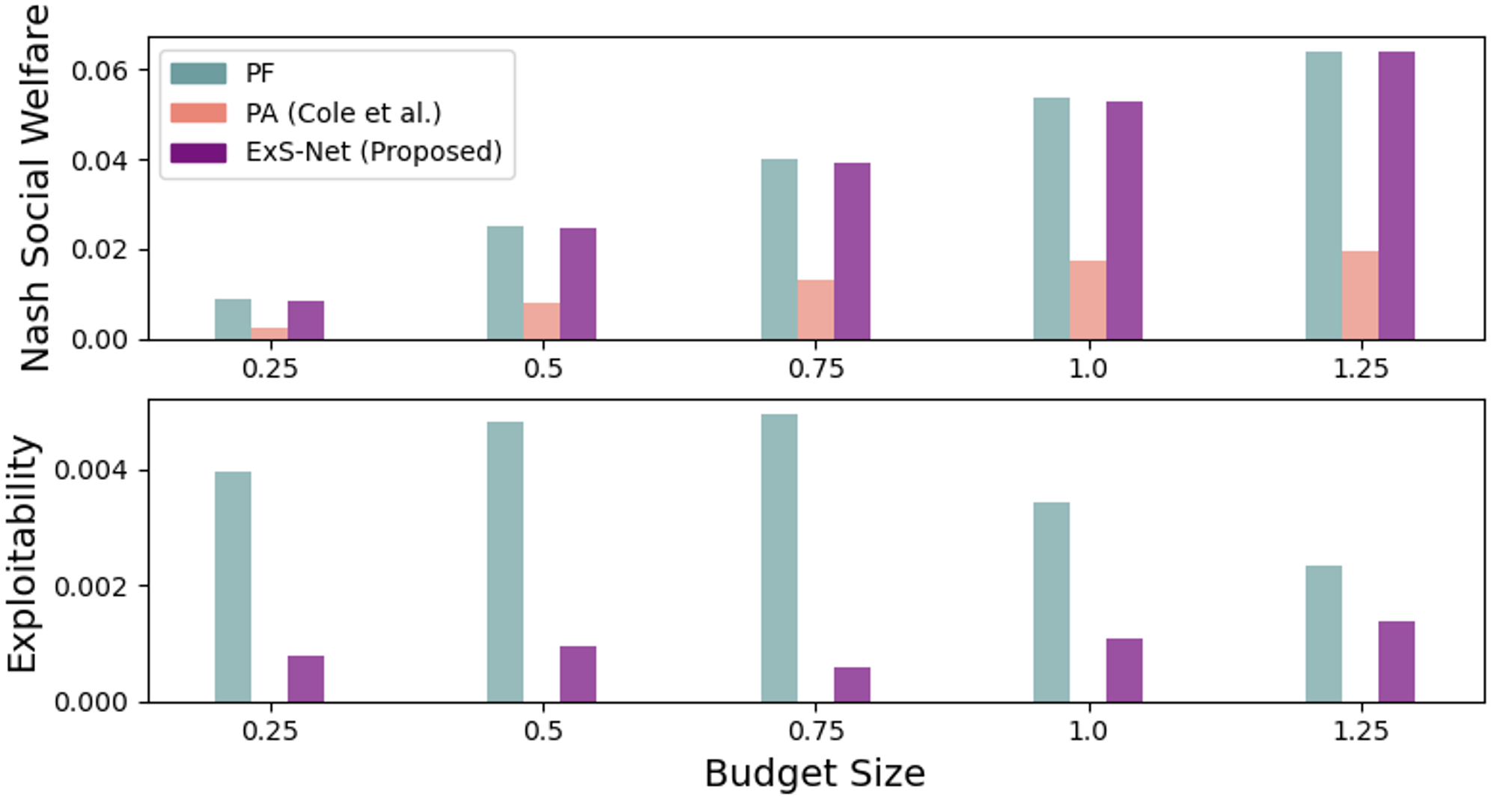}
  \caption{Performance of ExS-Net under varying budgets}
  % \vspace{-10pt}
  \label{fig:budget_change}
\end{figure}

We also investigate the impact of varying budget sizes. 
In Figure~\ref{fig:budget_change}, we plot of the NSW and exploitability of the mechanisms in a 2-agent 2-resource allocation problem where the budget for each resource increases from 0.25 (fierce competition) to 1.5 (little competition).
Figure~\ref{fig:budget_change} shows that ExS-Net consistently performs well across budget levels.\footnote{The tolerance $\epsilon$ on exploitability is constantly set to 1e-3 here. Setting $\epsilon$ with respect to the exploitability of PF across different budgets can further reduce the exploitability of ExS-Net without significant compromise in NSW when budget is 1 and 1.25.}

\subsection{Supplementary Material for Results in Section~\ref{sec:exp:dist_mismatch}}\label{sec:supp:dist_mismatch}
The randomly perturbed training samples are generated according to
\begin{align}
\begin{aligned}
    &\bar{v}_{i,m}^l \sim \text{Unif}(0.1,1),\,\, \bar{x}_{i,m}^l \sim \text{Unif}(0.1,1).\\
    &\tilde{v}_{i,m}^l\sim\text{Cauchy}(0,0.01), \,\,\tilde{x}_{i,m}^l\sim\text{Cauchy}(0,0.01),\\
    &v_{i,m}^l=[\bar{v}_{i,m}^l+\tilde{v}_{i,m}^l]_{[0.1,1]},\\
    &\widehat{x}_{i,m}\!\sim\!\text{Bern}(0.5), \, x_{i,m}^l=\widehat{x}_{i,m}[\bar{x}_{i,m}^l+\tilde{x}_{i,m}^l]_{[0.1,1]}.
\end{aligned}
\label{eq:dist_mismatch_random}
\end{align}

\begin{figure}[!ht]
  \centering
  \vspace{-2pt}
  \includegraphics[width=.5\linewidth]{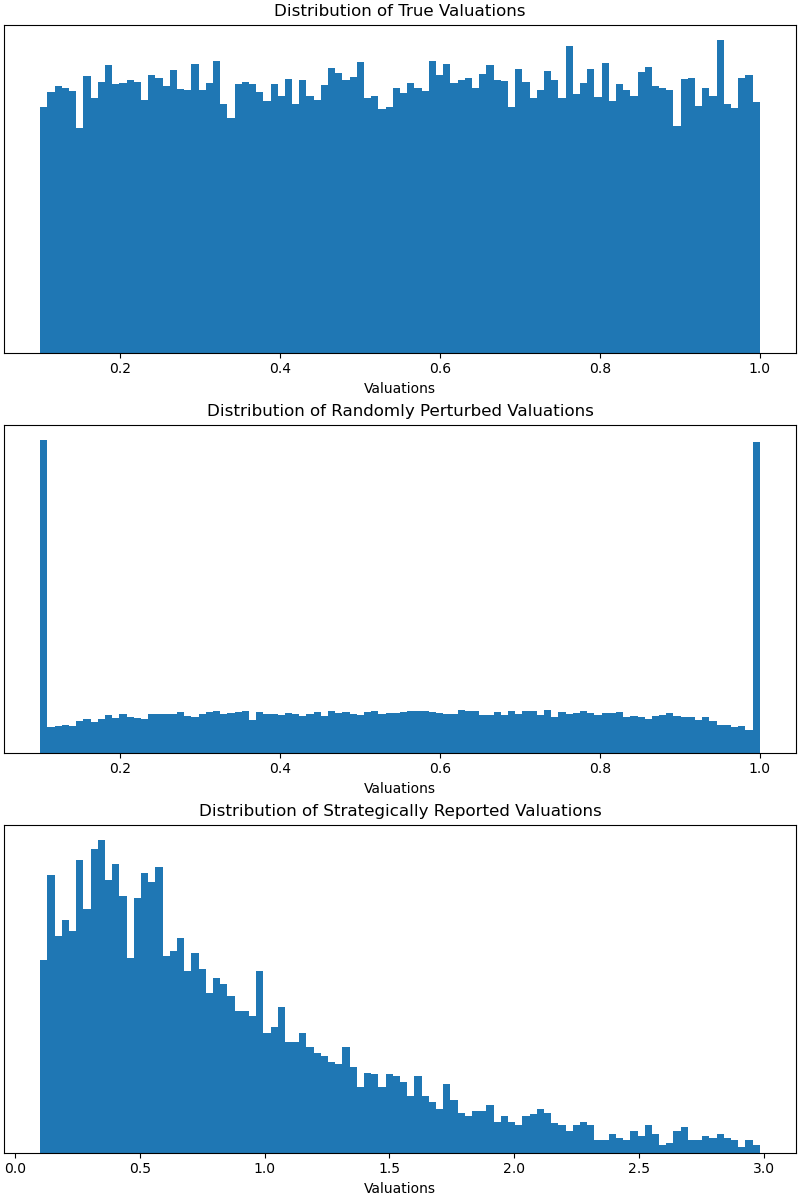}
  \caption{Distributions of untruthful valuations}
  \label{fig:misreport_dist}
\end{figure}

Figure~\ref{fig:misreport_dist} shows the probability density of the distributions described in \eqref{eq:dist_mismatch_random} and \eqref{eq:dist_mismatch_adversarial}.

\subsection{Simulation Details - Figure.~\ref{fig:exploitability_2x2}}\label{sec:fig:exploitability_2x2}

Figure.~\ref{fig:exploitability_2x2} is produced on a two-agent two-resource $(N=2,M=2)$ allocation problem. We consider a scarce resource scenario where $x=\1_{NM}$, $b=\1_{M}$, and $w=\1_{N}$. Both agents have a preference on the first resource, with $v_1=\{1,1/2\}$ and $v_2=\{1,1/4\}$ (note that only the relative preference among resources determines the PF allocation; the absolute magnitude of values does not play a role due to the log in the objective). When agent 2 reports truthfully, Figure~\ref{fig:exploitability_2x2} plots the utility of agent 1 as it varies the reported preference ratio $v_{1,2}/v_{1,1}$ from 0.1 to 3.

\section{Generalization bounds for the learning mechanisms} \label{sec:G_bnd}

\subsection{Definitions $\&$ Preliminaries}
Let~$\mathcal{F}$ denote a class of bounded functions~$f: Z \rightarrow [-c,c]$ defined on an input space~$Z$ for some~$c>0$. Let~$D$ be a distribution over~$Z$ and let~$\mathcal{S} = \{z_1,z_2,\cdots,z_L\}$ be a sample drawn i.i.d from some distribution~$D$ over input space~$Z$. 

\begin{definition}
The capacity of the function class~$\mathcal{F}$ measured in terms of empirical Rademacher complexity on sample~$\mathcal{S}$ is defined as
\[
\hat{\mathcal{R}}(\mathcal{F}):= \frac{1}{L} \mathbb{E}_{\sigma} \Bigg[ \sup_{f \in \mathcal{F}} \sum_{z_i \in \mathcal{S}} \sigma_i f(z_i) \Bigg],
\]
where~$\sigma \in \{-1,1 \}^L$ and each $\sigma_i$ is drawn from a uniform distribution on~$\{-1,1 \}$.
\end{definition}

\begin{definition}
The covering number of a set~$\mathcal{M}$, denoted as~$\mathcal{N}_{\infty}(\mathcal{M},\epsilon)$, is the minimal number of balls of radius~$\epsilon$ (measured in the $l_{\infty,1}$ distance) needed to cover the set~$\mathcal{M}$. 
\end{definition}
For example, the~$l_{\infty,1}$ distance between mechanisms~$f,f' \in \mathcal{M}$ is given as
\[
\max_{(v,x,b)} \sum_{i=1}^N \sum_{j=1}^M | f_{ij}(v,x,b) - f'_{ij}(v,x,b)|.
\]

\begin{lem}[\shortciteN{SSB14}] \label{lem:ERM}
Then with probability at least~$1-\delta$ over draw of~$S$ from~$D$, for~$f\in\mathcal{F}$,
\[
\mathbb{E}_{z \sim D} [f(z)] \leq \frac{1}{L} \sum_{l=1}^L f(z_l) + 2 \hat{\mathcal{R}}_L(\mathcal{F}) + 4c \sqrt{\frac{2 \log(4/\delta)}{L}}
\]
\end{lem}

\begin{lem} [Massart] \label{lem:Mas}
    Let~$\mathcal{G}$ be some finite subset of~$\mathbb{R}^m$ and $\sigma_1,\sigma_2,\cdots,\sigma_m$ be independent Rademacher random variables. Then,
    \[
    \mathbb{E} \Bigg[\sup_{g \in \mathcal{G}} \frac{1}{m} \sum_{i=1}^m \sigma_i g_i \Bigg] \leq \frac{ \sqrt{2 \Big(\sup_g \sum_i g_i^2 \Big) \log |G|}}{m}.
    \]
\end{lem}
Let~$\phi: \mathbb{R}^N \mapsto \mathbb{R}^N$ represent the activation function of the any layer for input~$s \in \mathbb{R}^{NM}$, given as 
\[
\phi = [\texttt{softmax}(s_{1,1},\cdots,s_{N,1}),\cdots,\texttt{softmax}(s_{1,m},\cdots,s_{N,M})],
\]
where~$\texttt{softmax}:\mathbb{R}^N \mapsto [0,1]^N$. Note that for any~$u \in \mathbb{R}^N$, 
\[
\texttt{softmax}_i(u) = \frac{e^{u_i}}{\sum_{k=1}^N e^{u_k}}.
\]
\begin{lem}[\shortciteN{dütting2022optimal}]  \label{lem:softmax}
For any~$s,s' \in \mathbb{R}^{NM}$, the activation function for a \texttt{softmax}~layer is $1-$Lipschitz, i.e.,
\[
\|\phi(s) - \phi(s')\|_1 \leq \|s-s' \|.
\]
\end{lem}

\begin{lem}[\shortciteN{dütting2022optimal}] \label{lem:cov_num}
    Let~$\mathcal{F}_k$ be a class of feed-forward neural networks that maps an input vector~$y \in \mathbb{R}^{d_0}$ to an output vector~$\mathbb{R}^{d_k}$, with each layer~$l$ containing~$T_l$ nodes and computing~$z \mapsto \phi_l(w^l z)$ and $\phi_l: \mathbb{R}^{T_l} \rightarrow [\psi, \psi]^{T_l}$. Further, for each network in~$\mathcal{F}_k$, let the parameter matrices~$\|w^l\|_1 \leq W$ and $\|\phi(s) - \phi(s') \| \leq \Phi \|s -s' \|_1$ for any~$s,s' \in \mathbb{R}^{T_{l-1}}$. The covering number of the network is
    \[
    \mathcal{N}_{\infty}(\mathcal{F}_k,\epsilon) \leq \Big \lceil \frac{2 \psi d^2 W (2\Phi W)^k}{\epsilon}\Big \rceil^d, 
    \]
    where~$T = \max_{l \in [k]} T_l$ and~$d$ is the total number of parameters in the network.
\end{lem}

\subsection{Proof of Theorem}

Consider a parametric class of mechanisms,~$f^{\omega} \in \mathcal{M}$, defined using parameters~$\omega \in \mathbb{R}^d$ for~$d>0$. With a slight abuse of notation, let~$u^{\omega}(v,x,b):= u(f^{\omega}(v,x,b),v,x)$ and let~$y \mapsto (v,x,b)$, $y_i^{'} \mapsto ((v_i',v_{-i}),(x_i',x_{-i}),b)$.  Consider the following function classes
% \begin{align*}
%     \mathcal{U}_i &= \Big\{u^{\omega}_i: Y \mapsto \mathbb{R} ~|~ u^{\omega}_i(y) = \sum_m v_{i,m} \min \{f_{i,m}^{\omega}(y), x_{i,m} \},~~\text{for some}~f^{\omega} \in \mathcal{M} \Big\}  \\
%     \text{NSW} \circ \mathcal{M} &= \Big\{ f: Y \mapsto \mathbb{R} ~|~ f(y) = \sum_{i=1}^N \log u_{i}^{\omega}(y),~~\text{for}~u^{\omega}_i \in \mathcal{U}_i \Big\} \\
%     \text{exp} \circ \mathcal{U}_i &= \Big\{ e_i: Y \mapsto \mathbb{R} ~|~ e_i(y) =  \max_{y_i^{'}}u_i^{\omega}(y_i^{'}) -u^{\omega}_i(y), ~~\text{for}~u^{\omega}_i \in \mathcal{U}_i \Big\} \\
%     \overline{\text{exp}} \circ \mathcal{U} &= \Big\{h: Y \mapsto \mathbb{R}~|~ h(y) = \sum_{i=1}^N e_i(y),~~\text{for some}~(e_1,e_2,\cdots,e_N) \in \text{exp} \circ \mathcal{U} \Big\}
% \end{align*}
\begin{align*}
    &\mathcal{U}_i=\hspace{-2pt} \big\{u^{\omega}_i\hspace{-2pt}:\hspace{-2pt} Y\hspace{-1pt} \mapsto \hspace{-1pt}\mathbb{R} | u^{\omega}_i(y) \hspace{-2pt}=\hspace{-3pt} \sum_m v_{i,m} \min \{f_{i,m}^{\omega}(y)\hspace{-1pt}, \hspace{-1pt}x_{i,m} \}~\text{for some}~f^{\omega}\hspace{-3pt} \in\hspace{-3pt} \mathcal{M} \big\}  \\
    &\text{NSW} \circ \mathcal{M} = \big\{ f: Y \mapsto \mathbb{R} ~|~ f(y) = \sum_{i=1}^N \log u_{i}^{\omega}(y)~\text{for}~u^{\omega}_i \in \mathcal{U}_i \big\} \\
    &\text{exp} \circ \mathcal{U}_i \hspace{-2pt}=\hspace{-2pt} \big\{ e_i: Y \mapsto \mathbb{R} ~|~ e_i(y) \hspace{-2pt}=\hspace{-2pt}  \max_{y_i^{'}}u_i^{\omega}(y_i^{'}) \hspace{-2pt}-\hspace{-2pt}u^{\omega}_i(y)~\text{for}~u^{\omega}_i \in \mathcal{U}_i \big\} \\
    &\overline{\text{exp}} \circ \mathcal{U} =\hspace{-2pt} \big\{h: Y \hspace{-2pt}\mapsto \hspace{-1pt}\mathbb{R}| h(y)\hspace{-2pt} =\hspace{-2pt} \sum_{i=1}^N e_i(y)~\text{for some}~(e_1,e_2,\cdots,e_N)\hspace{-2pt} \in\hspace{-2pt} \text{exp} \circ \mathcal{U} \big\}.
\end{align*}

\begin{proposition} \label{thm:RC_nsw}
For each agent~$i$, assume that~$\frac{1}{\psi} \leq  u^{\omega}_i(y) \leq \psi ,~\forall~y \in F$ and some $\psi>0$, and $v_i(S) \leq 1,~\forall~i$ and all subsets of resources. Let~$\mathcal{M}$ denote the class of mechanisms and fix~$\delta \in (0,1)$. With probability at least~$1-\delta$, over draw of~$L$ profiles from~$F$, for any parameterized allocation~$f^{\omega} \in \mathcal{M}$,
\begin{align*}
\mathbb{E}_{y \sim F}\Big[\sum_{i=1}^N \log u_i^{\omega}(y) \Big] \geq \frac{1}{L} \sum_{l=1}^L \sum_{i=1}^N \log u_i^{\omega}(y^{(l)}) - 2N \Delta_L - CN \sqrt{\frac{\log(1/\delta)}{L}},
\end{align*}
where~$C$ is a distribution independent constant and 
\[
\Delta_L = \inf_{\epsilon > 0} \Bigg\{ \epsilon + N(\log\psi+1)  \sqrt{\frac{2 \log(\mathcal{N}_{\infty}(\mathcal{M},\frac{\epsilon}{\psi}))}{L}} \Bigg\}.
\] 
\end{proposition}

\begin{proof}
Using Lemma~\ref{lem:ERM}, the result follows except the characterization of the empirical Rademacher complexity that we derive below. By the definition of the covering number, we have that for any~$h(y) \in \text{NSW} \circ \mathcal{M}$, there is a $\hat{h}(y) \in \widehat{\text{NSW}}\circ \mathcal{M}$ such that $\max_y |h(y) - \hat{h}(y)| \leq \epsilon$. We have the following
\begin{align*}
    \hat{\mathcal{R}}_L(\text{NSW} \circ \mathcal{M}) &= \frac{1}{L} \mathbb{E}_{\sigma} \Big[\sup_{u} \sum_{l=1}^L \sigma_l \sum_{i=1}^N \log u^{\omega}_i(y^{(l)}) \Big] \\
    &= \frac{1}{L} \mathbb{E}_{\sigma} \Big[\sup_{h} \sum_{l=1}^L \sigma_l \hat{h}(y^{(l)}) \Big] \hspace{-2pt}+\hspace{-2pt} \frac{1}{L} \mathbb{E}_{\sigma} \Big[\sup_{u} \sum_{l=1}^L \sigma_l \Big\{h(y^{(l)})  \hspace{-2pt}- \hspace{-2pt}\hat{h}(y^{(l)})\Big\} \Big] \\
    &\leq  \frac{1}{L} \mathbb{E}_{\sigma} \Big[\sup_{\hat{h}} \sum_{l=1}^L \sigma_l  \hat{h}(y^{(l)}) \Big] + \frac{1}{L} \mathbb{E}_{\sigma} \|\sigma \| \epsilon 
\end{align*}
The result follows from the following arguments:    
\begin{itemize}
    \item[i.)] We shall first establish that
\[
\mathcal{N}_{\infty}(\text{NSW} \circ \mathcal{M}) \leq \mathcal{N}_{\infty}(\mathcal{M},\frac{\epsilon}{\psi}).
\]
By definition of the covering number for the mechanism class~$\mathcal{M}$, there exists a cover~$\hat{\mathcal{M}}$ of size~$|\hat{\mathcal{M}}| \leq \mathcal{N}_{\infty}(\mathcal{M},\frac{\epsilon}{\psi})$ such that for any~$f^{\omega} \in \mathcal{M}$ there is a $\hat{f}^{\omega} \in \hat{\mathcal{M}}$ such that for all~$y$,
\[
\sum_{i,m} |f^{\omega}_{i,m}(y) - \hat{f}^{\omega}_{i,m}(y)| \leq \frac{\epsilon}{\psi}
\]
For~$g(y) = \sum_{i=1}^N \log u_{i}^{\omega}(y)$, we have  
\begin{align*}
    \Big|g(y) - \hat{g}(y)\Big|&= \Big|\sum_{i=1}^N \Big\{\log u_{i}^{\omega}(y) - \log \hat{u}_{i}^{\omega}(y) \Big\} \Big| \\
    &\hspace{10pt}\leq \psi \Big|\sum_{i=1}^N u_{i}^{\omega}(y) - \hat{u}_{i}^{\omega}(y) \Big| \\
    &\hspace{10pt}\leq \psi \Big| \sum_i \sum_m v_{i,m} \Big\{ \min \{f_{i,m}^{\omega}(y), x_{i,m} \} - \min \{\hat{f}_{i,m}^{\omega}(y), x_{i,m} \} \Big\} \Big| \\
    &\hspace{10pt}\leq \psi \sum_i \sum_m v_{i,m} \Big|\min \{f_{i,m}^{\omega}(y), x_{i,m} \} - \min \{\hat{f}_{i,m}^{\omega}(y), x_{i,m} \} \Big| \\
    &\hspace{10pt}\leq \psi \sum_i \sum_m v_{i,m} \Big|f_{i,m}^{\omega}(y) -  \hat{f}_{i,m}^{\omega}(y) \Big| < \epsilon.
\end{align*}
\item[ii.)] We have from Massart's lemma (Lemma~\ref{lem:Mas}),
\[
\hat{\mathcal{R}}_L(\text{NSW} \circ \mathcal{M}) \leq \sqrt{\sum_l \Big(\hat{h}(y^{(l)})\Big)^2}  \frac{\sqrt{2 \log (\mathcal{N}_{\infty}(\text{NSW} \circ \mathcal{M}),\epsilon)}}{L} + \epsilon.
\]
\item[iii.)] We have the trivial bound,
\[
\sqrt{\sum_l \Big(\hat{h}(y^{(l)})\Big)^2} \hspace{-2pt}\leq\hspace{-2pt} \sqrt{\sum_l \Big(\sum_i \log u^{\omega}_i (v^{(l)}) + n\epsilon\Big)^2} \hspace{-2pt}\leq \hspace{-2pt}N(\log\psi+1) \sqrt{L}.
\]
\end{itemize}
The result follows.  
\end{proof}

\begin{proposition}
Let $\text{exp}_i(\omega) := \mathbb{E}_{y} \Big[ \max_{y'}u^{\omega}_i(y_i^{'})-u_i^{\omega}(y)\Big]$ and $\widehat{\text{exp}}_i(\omega) := \frac{1}{L} \sum_{l=1}^L \Big[ \max_{\bar{y}}u^{\omega}_i(\bar{y}_i^{(l)})-u_i^{\omega}(y^{(l)})\Big]$. Under the same assumptions as in Theorem~\ref{thm:RC_nsw}, the empirical exploitability satisfies the following:
    \[
    \frac{1}{N} \sum_{i=1}^N \text{exp}_i(\omega) \leq  \frac{1}{N} \sum_{i=1}^N \widehat{\text{exp}}_i(\omega) + \Delta^{e}_L + C' \sqrt{\log(1/\delta)/L},
    \]
   where~$C$ is a distribution independent constant and 
\[
\Delta^{e}_L = \inf_{\epsilon > 0} \Bigg\{ \epsilon + (2\psi+1)N  \sqrt{\frac{2 \log(\mathcal{N}_{\infty}(\mathcal{M},\frac{\epsilon}{2N}))}{L}} \Bigg\}.
\]
\end{proposition}

\begin{proof}
As before, using Lemma~\ref{lem:ERM}, the result follows except the characterization of the empirical Rademacher complexity of the class~$\hat{\mathcal{R}}(\overline{\text{exp}} \circ \mathcal{U})$, which we do below. The proof builds on the following results.
\begin{itemize}
    \item[i.)] $\mathcal{N}_{\infty}(\text{exp} \circ \mathcal{U}_i,\epsilon) \leq \mathcal{N}_{\infty}(\mathcal{U}_i,\frac{\epsilon}{2})$. \\
    By the definition of covering number~$\mathcal{N}_{\infty}(\mathcal{U}_i,\epsilon)$, there exists a cover~$\hat{\mathcal{U}}_i$ with size at most~$\mathcal{N}_{\infty}(\mathcal{U}_i,\frac{\epsilon}{2})$ such that for any~$u^{\omega}_i \in \mathcal{U}_i$ there is a~$\hat{u}^{\omega}_i \in \hat{\mathcal{U}}_i$ with
    \[
    \max_{y} |u^{\omega}_i (y) -  \hat{u}^{\omega}_i(y) | \leq \frac{\epsilon}{2}.
    \]
    We have for the exploitability for each agent~$i$ with any~$y$,
    \begin{align*}
        &|\max_{y_i^{'}}u_i^{\omega}(y_i^{'}) -u^{\omega}_i(y) - \max_{y_i^{'}}\hat{u}_i^{\omega}(y_i^{'}) +\hat{u}^{\omega}_i(y)| \\
        &\hspace{20pt}\leq |\max_{y_i^{'}}u_i^{\omega}(y_i^{'}) - \max_{y_i^{'}}\hat{u}_i^{\omega}(y_i^{'})| + |u^{\omega}_i(y) - \hat{u}^{\omega}_i(y)| \leq \epsilon.
    \end{align*}
where the last inequality follows from the following relation.
\begin{align*}
&\max_{\bar{y}_i}u_i^{\omega}(\bar{y}_i) = u_i^{\omega}(y^*) \\
&\hspace{20pt}\leq \hat{u}_i^{\omega}(y^*) + \epsilon/2 \leq \hat{u}_i^{\omega}(\bar{y}^*) + \epsilon/2 \leq \max_{\bar{y}} \hat{u}_i^{\omega}(\bar{y}) + \epsilon,\\
&\max_{y_i}\hat{u}_i^{\omega}(y_i) = \hat{u}_i^{\omega}(y^*) \\
&\hspace{20pt}\leq u_i^{\omega}(y^*) + \epsilon/2 \leq u_i^{\omega}(\bar{y}^*) + \epsilon/2 = \max_{\bar{y}} u_i^{\omega}(\bar{y}) + \frac{\epsilon}{2}.
\end{align*}
\item[ii.)] $\mathcal{N}_{\infty}(\overline{\text{exp}} \circ \mathcal{U},\epsilon) \leq \mathcal{N}_{\infty}(\mathcal{U},\frac{\epsilon}{2N}).$
The result following by considering a cover~$\mathcal{N}_{\infty}(\mathcal{U},\frac{\epsilon}{2N})$ such that
\[
    \max_{y} \sum_{i=1}^N |u^{\omega}_i (y) -  \hat{u}^{\omega}_i(y) | \leq \frac{\epsilon}{2N},
\]
and using the same arguments as above.
\item[iii.)] $\mathcal{N}_{\infty}(\mathcal{U},\epsilon) \leq \mathcal{N}_{\infty}(\mathcal{M},\epsilon)$.\\
By definition of the covering number for the mechanism class~$\mathcal{M}$, there exists a cover~$\hat{\mathcal{M}}$ of size~$\hat{\mathcal{M}} \leq \mathcal{N}_{\infty}(\mathcal{M},\epsilon)$ such that for any~$f^{\omega} \in \mathcal{M}$ there is a $\hat{f}^{\omega} \in \hat{\mathcal{M}}$ such that for all~$y$,
\[
\sum_{i,m} |f^{\omega}_{i,m}(y) - \hat{f}^{\omega}_{i,m}(y)| \leq \epsilon.
\]
We have for the class~$\mathcal{U}$,
\begin{align*}
    \Big|\sum_{i=1}^N u_{i}^{\omega}(y) - \hat{u}_{i}^{\omega}(y) \Big| 
    &\leq \Big| \sum_i \sum_m v_{i,m} \Big\{ \min \{f_{i,m}^{\omega}(y), x_{i,m} \} - \min \{\hat{f}_{i,m}^{\omega}(y), x_{i,m} \} \Big\} \Big| \\
    &\leq  \sum_i \sum_m v_{i,m} \Big|\min \{f_{i,m}^{\omega}(y), x_{i,m} \} - \min \{\hat{f}_{i,m}^{\omega}(y), x_{i,m} \} \Big| \\
    &\leq \sum_i \sum_m v_{i,m} \Big|f_{i,m}^{\omega}(y) -  \hat{f}_{i,m}^{\omega}(y) \Big| < \epsilon.
\end{align*}
\end{itemize}
The result follows from Lemma~\ref{lem:Mas} and similar  arguments as in Proposition~\ref{thm:RC_nsw}.
\end{proof}

\begin{theorem}[Generalization Bound]
Consider ExS-Net $f^{\omega}$ mechanisms parameterized by a neural network with~$R$ hidden layers,~$K$ nodes per hidden layer, ReLU activation, a total of $d$ parameters, and the vector of all model parameters $\| \omega\|_1 \leq \Omega$. With probability at least~$1-\delta$, 
\begin{align*}
\max\{\varepsilon_{\logNSW}(f^{\omega},L),\varepsilon_{\mathbf{exp},i}(f^{\omega},L)\}\leq \Ocal\Big(\!\frac{\sqrt{Rd \log(\! LN\Omega \max\{K,MN\})}}{L}\!+N\sqrt{\frac{\log(1/\delta)}{L}}\,\Big)\hspace{-1pt}.
\end{align*}
\end{theorem}
\begin{proof}
    First, note that the activation functions on all hidden layers are ReLU functions, which are 1-Lipschitz. The output layer of ExS-Net is $1-$Lipschitz from Lemma~\ref{lem:softmax}. Using the definitions of~$\Delta$'s with~$\Phi = 1, d = \max\{K, MN\}$ in Lemma~\ref{lem:cov_num}, the result follows using Proposition~1 and Proposition~2.
\end{proof}

%%%%%%%%%%%%%%%%%%%%%%%%%%%%%%%%%%%%%%%%%%%%%%%%%
%%%%%%%%%%%%%%%%%%%%%%%%%%%%%%%%%%%%%%%%%%%%%%%%%
%%%%%%%%%%%%%%%%%%%%%%%%%%%%%%%%%%%%%%%%%%%%%%%%%
%%%%%%%%%%%%%%%%%%%%%%%%%%%%%%%%%%%%%%%%%%%%%%%%%

% \input{Appendix_DistributionMismatch}

\end{document}